\documentclass[10pt,journal,compsoc]{IEEEtran} 

\ifCLASSOPTIONcompsoc
  \usepackage[nocompress]{cite}
\else
  \usepackage{cite}
\fi
\usepackage{mathrsfs}

\usepackage{makecell}
\usepackage{cellspace}
\usepackage{subfigure}

\newcommand\content{\textbf{\textit{Adaptive Content}}}
\newcommand\assessment{\textbf{\textit{Adaptive Assessment}}}
\newcommand\sequence{\textbf{\textit{Adaptive Sequence}}}
\newcommand\model{PAL}
\newcommand\system{MOOC-Guide}
\newcommand\dataset{MOOC2.7M}
\usepackage{fancyhdr}
\usepackage{multirow}
\usepackage{balance}
\usepackage[utf8]{inputenc} %
\usepackage[T1]{fontenc}    %
\usepackage{xcolor}	
\usepackage{times,flushend}
\usepackage{amsmath}
\usepackage{amsthm}
\usepackage{amsfonts}
\usepackage{amssymb}
\usepackage{nicefrac}
\usepackage{textcomp}
\usepackage{mathrsfs}
\usepackage{makecell}
\usepackage{cellspace}
\usepackage{bm}
\usepackage{multirow}
\usepackage{multicol}
\usepackage{hyperref}
\usepackage{cleveref}
\usepackage{graphicx}
\graphicspath{{figures/}}
\usepackage{booktabs}
\usepackage{color}
\usepackage{tabularx}
\usepackage{threeparttable}
\usepackage{algorithm}
\usepackage{algorithmicx}
\usepackage{algpseudocode}

\include{symbols}

\begin{document}

\title{Towards a General Pre-training Framework for Adaptive Learning in MOOCs}

\author{Qingyang~Zhong*,~Jifan~Yu*,~Zheyuan~Zhang,~Yiming~Mao,~Yuquan~Wang,~Yankai~Lin,\\Lei~Hou,~Juanzi~Li,~\IEEEmembership{Member,~IEEE},~Jie~Tang,~\IEEEmembership{Fellow,~IEEE}%
\IEEEcompsocitemizethanks{
\IEEEcompsocthanksitem * Equal Contribution. Jie Tang is the corresponding author.\protect
\IEEEcompsocthanksitem Qingyang Zhong, Jifan Yu, Zheyuan Zhang and Yuquan Wang are with the Department of Computer Science and Technology, Tsinghua University, China. E-mail: \{zqy20, yujf18, zheyuan-18, wangyuqu19\} @mails.tsinghua.edu.cn. This work was done when Yiming Mao was at Tsinghua University. E-mail: 845285227@qq.com.\protect
\IEEEcompsocthanksitem Yankai Lin is with Pattern Recognition Center, WeChat AI, Tencent Inc. E-mail: yankailin@tencent.com.%
\IEEEcompsocthanksitem Lei~Hou, Juanzi~Li and Jie~Tang are with the Department of Computer Science and Technology, Tsinghua University, China. E-mail: \{houlei, lijuanzi, jietang\}@tsinghua.edu.cn\protect}%
}

\IEEEtitleabstractindextext{%
\begin{abstract}
Adaptive learning aims to stimulate and meet the needs of individual learners, which requires sophisticated system-level coordination of diverse tasks, including modeling learning resources, estimating student states, and making personalized recommendations. Existing deep learning methods have achieved great success over statistical models; however, they still lack generalization for diverse tasks and suffer from insufficient capacity since they are composed of highly-coupled task-specific architectures and rely on small-scale, coarse-grained recommendation scenarios. To realize the idea of general adaptive systems proposed in pedagogical theory, with the emerging pre-training techniques in NLP, we try to conduct a practical exploration on applying pre-training to adaptive learning, to propose a unified framework based on data observation and learning style analysis, properly leveraging heterogeneous learning elements. Through a series of downstream tasks of \emph{Learning Recommendation}, \emph{Learning Resource Evaluation}, \emph{Knowledge Tracing}, and \emph{Dropout Prediction}, we find that course structures, text, and knowledge are helpful for modeling and inherently coherent to student non-sequential learning behaviors and that indirectly relevant information included in the pre-training foundation can be shared across downstream tasks to facilitate effectiveness. We finally build a simplified systematic application of adaptive learning and reflect on the insights brought back to pedagogy. The source code and dataset will be released. %
\end{abstract}

}

\maketitle
\thispagestyle{fancy}
\renewcommand{\headrulewidth}{0pt}
\renewcommand{\footrulewidth}{0pt}
\fancyfoot[C]{This work has been submitted to the IEEE for possible publication. Copyright may be transferred without notice, after which this version may no longer be accessible.}

\IEEEdisplaynontitleabstractindextext

\IEEEpeerreviewmaketitle

\section{Introduction}
\label{sec:introduction}

\IEEEPARstart{T}{here} \emph{are no talented people in the world. The problem is that educators should discover each student’s endowment and expertise and provide sufficient conditions}, said Sukhomlinsky, the renowned humanistic educator of the twentieth century. Adaptive learning, also called adaptive teaching, is an educational method employing algorithms to orchestrate interaction and provide customized resources and learning activities to stimulate and meet the needs of individual learners~\cite{kaplan2021higher}, which consists of multiple tasks, including modeling learning resources~\cite{pan2017prerequisite, yu2019course}, estimating student mastery of skills~\cite{piech2015deep, chen2018prerequisite}, and making personalized learning recommendations~\cite{liu2019exploiting,bulathwela2020truelearn,tkde2021recommendation}. The prosperity of Massive Open Online Courses (MOOCs) has made adaptive learning increasingly crucial and greatly facilitated~\cite{yu2021mooccubex} since MOOCs attract large-scale learners of diversified backgrounds and knowledge levels, highly demanding tailored learning and targeted guidance that traditional teaching approaches cannot satisfy.

With the growing number of technical attempts~\cite{yu2019course,feng2019understanding,liu2019exploiting}, %
adaptive learning in MOOCs has entered the era of deep learning. Although these deep learning methods have outperformed traditional statistical models~\cite{ferguson2012learning}, they still suffer from two main problems. On the one hand, some~\cite{zhang2019hierarchical,liu2019exploiting} only target a single task and require optimization on abundant expert-annotated task-specific data, which cannot accumulate generalization across tasks. On the other hand, emerging approaches~\cite{bulathwela2020truelearn,tkde2021recommendation,gong2020sigir} are designed with sophisticated, heavily-engineered architectures to introduce multiple educational elements like student background and knowledge structures, which can only adapt to small, coarse-grained scenarios. Therefore, there is a need to explore more rational ways of exploiting large-scale, fine-grained data throughout the learning process provided by integrated educational datasets launched in recent years, such as Tutorialbank~\cite{fabbri2018tutorialbank} and MOOCCubeX~\cite{yu2021mooccubex}. %

In the face of these predicaments, pedagogical researchers have proposed \emph{Epistemic Frame Theory}~\cite{shaffer2016tutorial} to unify diverse adaptive learning tasks, calling for general adaptive systems~\cite{martin2020systematic}. Meanwhile, in computer technology, the explosion of pre-trained models in NLP, such as ELMO~\cite{Peters:2018}, BERT~\cite{devlin2019bert}, GPT~\cite{brown2020gpt3}, and T5~\cite{raffel2019exploring}, has profoundly changed the paradigm of deep learning and facilitated other AI research fields, offering an attractive tool for unifying varied tasks and realizing general systems. Therefore, we try to conduct a practical exploration to apply the pre-training technique to the scenario of adaptive learning by modeling student learning behaviors. Since student behaviors in MOOCs are sparse and fragmented and contain diversified information, we hope to analyze and answer the following three critical questions:  %

\begin{itemize}
    \item[\textbf{\textit{Q1}}] \textit{What elements in learning processes do contribute to the learning behavior modeling, and what are their roles?}
    \item[\textbf{\textit{Q2}}] \textit{Is indirectly relevant information contained in unified pre-training helpful across different adaptive learning tasks? }
    \item[\textbf{\textit{Q3}}] \textit{Do the findings from pre-training confirm or supplement the current pedagogical perspectives?}
\end{itemize}

With the assistance of MOOCCubeX~\cite{yu2021mooccubex}, which covers a wide range of learning resources and detailed student activities from XuetangX, the largest MOOC platform in China, we conduct a thorough learning style analysis in online learning environments. Through the analysis, we observe that watching videos is essential in MOOCs and that student learning behaviors are consistent with \emph{Felder-Silverman}'s~\cite{graf2007depth} non-sequential pattern but still obey underlying coherence, where the course structure, text, and knowledge background of each behavior reveal content relevance and global connections in student learning sequences. %

As student learning behaviors implicitly involve diversified educational elements, we propose a pre-training framework \model~for adaptive learning in place of heavily-engineered architectures to model students' non-strictly sequential behaviors employing a bidirectional Transformer~\cite{devlin2019bert} as the backbone. To unify different downstream tasks and benefit them from the pre-training stage, we introduce multi-level embedding layers to exploit heterogeneous learning elements considering text, knowledge concepts, and meta-information and design a specific `Masked Learning Behavior Prediction' task as the pre-training objective. %

Based on the pre-training foundation, we establish a series of benchmarks for various adaptive learning tasks in MOOCs to conduct comprehensive experimental exploration, including \emph{Learning Recommendation}~\cite{liu2019exploiting}, \emph{Learning Resource Evaluation}~\cite{ramesh2014learning}, \emph{Knowledge Tracing}~\cite{wang2020neural}, and \emph{Dropout Prediction}~\cite{feng2019understanding}. The experimental results show that the unified pre-training framework can better integrate diverse learning elements shared across downstream adaptive learning tasks, promoting effectiveness even if the information is indirectly relevant. The systematic application of resource retrieval and recommendation implemented on \model~can improve students learning outcomes aligned with pedagogy views.  %

In summary, our contributions are as follows:

$\bullet$ We thoroughly explore applying pre-training to adaptive learning and propose a unified framework to model student learning behaviors considering heterogeneous learning elements, including course structures, text, and knowledge. %

$\bullet$ We build a series of benchmarks for several adaptive learning tasks to evaluate and analyze the effectiveness of the proposed method, which is conceptually simple and empirically powerful, and find that indirectly relevant information considered in the pre-trained base can facilitate downstream performance.%

$\bullet$ We conduct a comprehensive survey and analysis of adaptive learning and provide an example of a systematic application to benefit the development of both pedagogy and machine learning for intelligent education. %

\section{Background}
\label{sec:background}

\subsection{Adaptive Learning}

Adaptive learning aims to address the unique needs of self-paced students by delivering custom learning experiences instead of one-size-fits-all learning patterns~\cite{kaplan2021higher}. With optimal learning path planning at its core, adaptive learning consists of multiple components, such as modeling learning resources, estimating students' mastery of skills, and making personalized learning recommendations, highly relevant to pedagogy theory, educational data mining, and learning analytics~\cite{bulathwela2020truelearn}.  It is widely accepted that adaptive learning can be categorized into three research directions~\cite{yu2021mooccubex}.

\content, the \emph{expert module}, focuses on the modeling, evaluation, and analysis of learning resources, whose typical tasks include education knowledge modeling~\cite{pan2017prerequisite, yu-etal-2020-mooccube} and automatic evaluation of materials~\cite{pardos2014affective}. It typically adopts the methods of natural language processing~\cite{pan2017course}, data mining~\cite{yu2019course} and knowledge modeling~\cite{fabbri2018tutorialbank} for resource understanding and comparison. %

\assessment, the \emph{student module}, comprises both micro and macro levels, with the former focusing on students' fine-grained behaviors on specific videos or questions and the latter on modeling the long-term traits. Micro-level attempts include cognitive modeling~\cite{wang2020neural} and knowledge tracing~\cite{piech2015deep,chen2018prerequisite}, which entails analyzing students' problem-solving records, while macro-level tasks apply to steady tendency prediction~\cite{feng2019understanding,dalipi2018mooc,moreno2020temporal}.

\sequence, the \emph{instruction module}, is a crucial topic in adaptive learning. It requires simultaneous modeling of learning resources and students and fully considering the students' historical performance, knowledge states, and candidates' content to recommend proper learning resources~\cite{bulathwela2020truelearn,liu2019exploiting}. Existing studies on this topic tend to relate it to sequence recommendation~\cite{pre4recstrategy} or knowledge structure acquisition~\cite{tkde2021recommendation}.

With the prosperity of deep learning, neural networks are widely employed in diverse adaptive learning approaches. Despite their significant improvement over traditional statistical methods, these efforts still rely on expert-annotated coarse-grained data, which makes them suffer from low transferability and insufficient capacity. Recently, researchers have summarized the existing methods and identified that the urgent future of adaptive learning research is to construct a systematical adaptive learning framework~\cite{martin2020systematic}, which inspires a few pioneers to explore how to integrate different components of adaptive learning~\cite{liu2019exploiting,yu2021mooccubex} to build general adaptive systems based on pedagogy insights~\cite{shaffer2017epistemic}. %

\subsection{Pretrained Models}

The explosion of pre-trained models in NLP, such as ELMo~\cite{Peters:2018}, BERT~\cite{devlin2019bert}, GPT~\cite{brown2020gpt3} and T5~\cite{raffel2019exploring}, has profoundly changed the whole paradigm of deep learning and gradually extended its influence to other AI research fields. The pre-training model originates from language modeling and performs self-supervised training on a large-scale corpus to achieve better initialization before proceeding with downstream tasks. The pre-training-fine-tuning paradigm has rapidly achieved SOTA results in various downstream NLP tasks with excellent generalization ability~\cite{brown2020gpt3} benefiting from the modeling of the language pattern, even in low-resource scenarios. This idea motivates research on conducting multimodal pretraining in other areas, including images (ViT~\cite{dosovitskiy2020image}), figure-text tasks (Dalle~\cite{ramesh2021zero}, CogView~\cite{ding2021cogview}), and large-scale graphs (GCC~\cite{qiu2020gcc}). Although some NLP researchers have trained language models specifically on education-related texts, such models are limited in assisting \content~tasks while ignoring crucial behavioral information of students. We argue that pre-training models should be designed with student behaviors to facilitate various adaptive downstream tasks for improving performance, especially in weak-resource scenarios. With the ability to integrate more dimensions of learning features, such a pre-training framework can become the foundation of a general adaptive system.

\section{Data Analysis}
\label{sec:data_analysis}
XuetangX\footnote{\url{https://next.xuetangx.com}}, launched in 2013, is now one of the largest MOOC platforms in China. It provides a wealth of learning records and has spawned many educational dataset efforts~\cite{yu-etal-2020-mooccube, yu2021mooccubex}. With the support of MOOCCubeX~\cite{yu2021mooccubex}, we build on its foundation to collect and process the learning logs of various types of students' behaviors for data statistics and perform learning style analysis to identify the critical elements of the learning process.

\subsection{Data Collection \& Processing}
\label{sec:data_processing}

When students study on XuetangX, the platform records multiple types of learning behaviors, such as watching videos (watch, stop, and jump), participating in forum discussion (comment and reply), and doing questions  (with correct/incorrect answers). The students' learning records in MOOCCubeX are from January 19, 2020, to November 3, 2020, containing 154,332,174 raw video-watching logs. However, these raw data are challenging to use directly since the video-watching raw data are `HeartBeat' logs at five-second intervals to record which video and relative position the student is currently watching. It is deemed necessary to aggregate the `Heartbeat' logs at proper granularity.

We sort the logs by time for each student and then aggregate the consecutive watching records of the same video, i.e., two neighbor records of the student are from the same video, and the current record is within $5$ seconds of the previous one. In this way, we convert the video-watching logs into $2,680,833$ fine-grained behavioral data of the students, including the watched video and duration. Since videos in MOOCs are widely equipped with subtitles corresponding to the playing position in milliseconds, we preserve the subtitles and link them to the students' behaviors to reflect the watched content. 

Besides, learning resources in MOOCs are not independent since the videos of a specific course are packed into several chapters and accompanied by exercises. Such structured information help organize the videos into a tree-like hierarchy. Therefore, we also take the course structures from MOOCCubeX to provide connections between learning resources for further research.

\subsection{Data Observation}
\label{sec:data_observation}

\subsubsection{Data Statistics}

\begin{table}[t]
\centering
\small
\setlength{\abovecaptionskip}{0.2cm}
\setlength{\belowcaptionskip}{0.cm}
\setlength{\tabcolsep}{1.2mm}
\caption{The statistics of the collected dataset \dataset.}%
\begin{tabular}{cc}
    \toprule
        Statistics & Number \\
    \midrule
        \# of courses & 4,124 \\
        \# of videos & 181,951 \\ %
        \# of questions & 1,026,303 \\
        \# of students & 117,992 \\
        \# of forum comments and replies & 15,719,235 \\
        \# of video-watching behaviors  & 2,680,833 \\
\bottomrule
\end{tabular}
\label{tab:behavior_statistics}
\end{table}

From MOOCCubeX, we find that the amount of various learning behaviors is exceptionally vast. Looking into each student's behaviors, we observe that almost all students have watched the videos at least once, while quite a few have no record of doing questions or participating in forum discussions. It makes sense that video-watching behaviors are common among students and across disciplines since the primary purpose of studying online is to gain knowledge by watching videos~\cite{roy2019inferring}. In this case, we take the video-watching behaviors as the essential for the learning process%
, and ultimately, we filter out students with less than five video-watching behaviors and fetch their corresponding question-doing and forum discussion behaviors from MOOCCubeX to build the dataset MOOC2.7M as shown in Table~\ref{tab:behavior_statistics}.

\subsubsection{Data Characteristics}

After aggregating the consecutive watching records, we denote each video-watching behavior as $x^{V_k}$, where $V_k$ denotes the unique VideoID of the watched video. Thus, a typical sequence of video-watching behaviors can be expressed as $\mathcal{X}^{\prime}=\left (x^{V_1}, x^{V_4}, x^{V_4}, x^{V_4}, x^{V_2}, x^{V_2}\right )$, where the student has interacted with video $V_1$, $V_4$, and $V_2$ chronologically,  $V_4$ three times and $V_2$ twice. The repeated behaviors usually come from two situations: \textbf{Stop} (pauses and continues to watch, resulting in a discontinuity of watching duration) and \textbf{Jump} (jumps forward or backward, resulting in a discontinuity of the watched content). We count the repeated times of neighbor behaviors in the learning sequences and find that the average is 1.6, while students act more actively in popular videos. Hence, repetition aggravates the behavior sparsity and increases the calculation burden. Since there is little change in the student's state of knowledge when watching the same video, it is reasonable to combine continuously repeated behaviors to alleviate such a dilemma. Therefore, we convert the above sequence to $\mathcal{X}=\left (x^{V_1}, x^{V_4}, x^{V_2}\right )$, and subsequently simply represent the learning sequences with time-ordered VideoIDs.

\begin{figure}[t]
    \centering
    \includegraphics[width=0.95\linewidth]{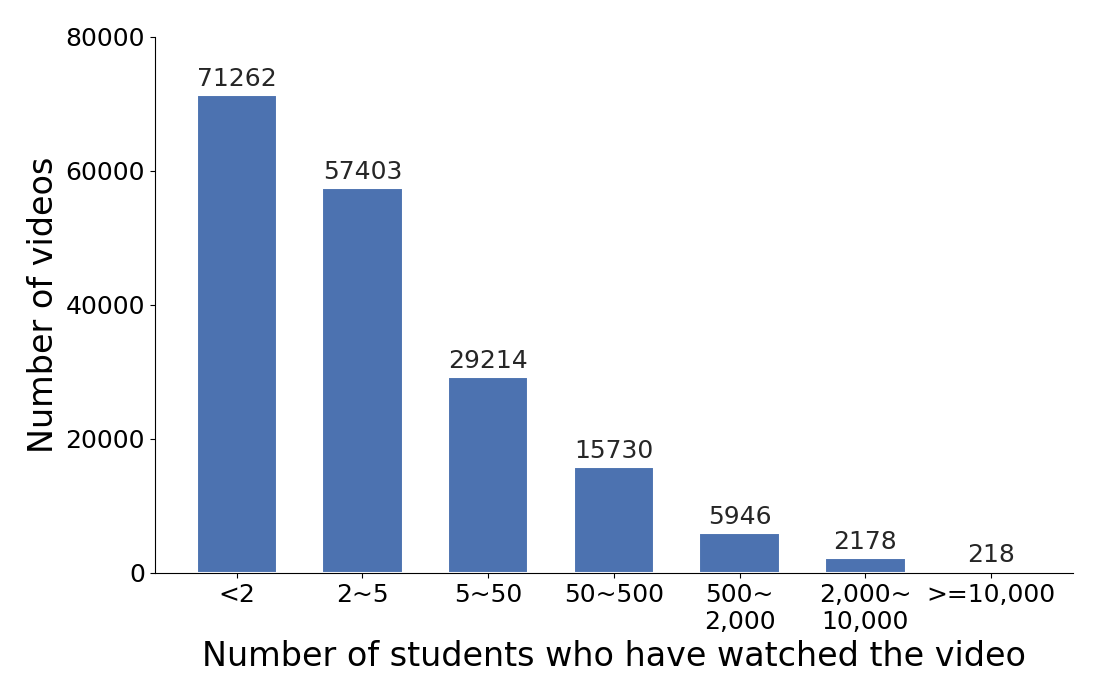}
    \caption{Long-tail distribution of video popularity.} 
    \label{fig:long_tail}
\end{figure}

To quantitatively demonstrate the sparsity of learning behaviors, we calculate the proportion of '0' (meaning 'not watched') in the student-video watching matrix. The ratio is surprisingly high at 99.88\%. For in-depth investigation, we divide videos into groups according to the number of students who have watched the video (Figure~\ref{fig:long_tail}) and find that 39.2\% of the videos have only been watched by one student, while 33,650 students have watched the most popular video, almost 28.5\% of the total. The results show extreme unevenness and long-tail distribution and that plenty of videos are involved in limited interactions. It is essential to conduct a learning style analysis on students' behaviors to mine critical learning components and alleviate such severe sparsity.

\subsection{Learning Style Analysis}
\label{sec:learning_style_analysis}

\begin{figure*}[t]
    \centering
    \includegraphics[width=0.95\textwidth]{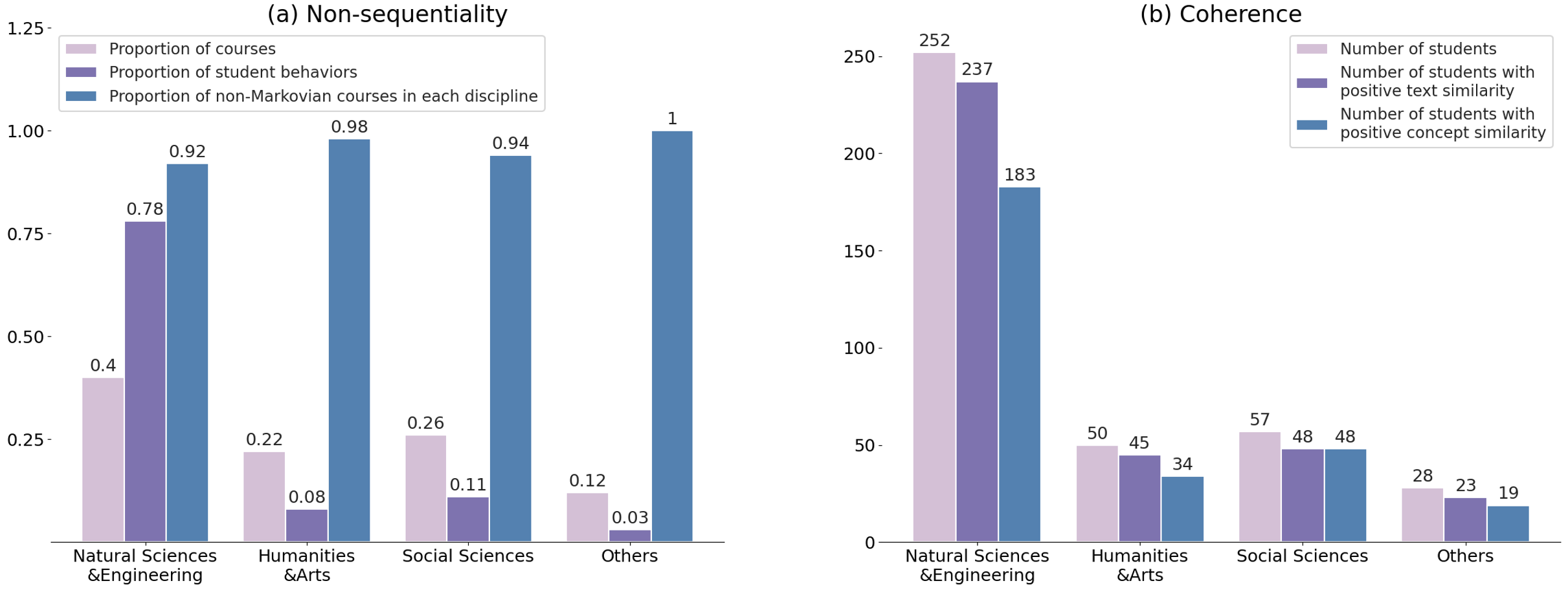}
    \caption{Learning analysis results. (a) shows the proportion of courses and student behaviors of each discipline and the proportion of non-Markovian courses in each discipline. (b) shows the number of students in total, those whose text similarity over the threshold, and those whose concept similarity over the threshold of each discipline.}
    \label{fig:data_analysis}
\end{figure*}

According to the \emph{Felder-Silverman Model}~\cite{graf2007depth}, a widely applied learning style scale that presents diverse dimensions of learning preferences, students can be either \emph{Sequential} or \emph{Global} learners. \emph{Sequential} learners prefer linear and logically ordered materials, whereas \emph{Global} learners prefer fragmented learning, revealing a non-sequential and coherent nature. We intend to explore the learning patterns of the students in the online environment.

\subsubsection{Sampling}
\label{sec:sampling}

First, we sample the students to represent the overall situation. The error of sampling inference can be calculated according to the sample size and controlled within the allowable range. Because the amount of students is enormous, we determine the sample size by $z^2p(1-p)/d^2$, where z is 1.96 for 95\% confidence, d is 0.05 for accuracy, and p is generally set to 0.5. Therefore, the accuracy of the survey results can be guaranteed as long as the sampling size is over 384. 
Finally, we randomly sampled 387 students from the total with 7,602 learning behaviors, and the videos they have watched are from 244 courses divided into four disciplines: Natural Sciences\&Engineering, Humanities\&Arts, Social Sciences, and Others. From the sample, we can see that most courses are from Natural Sciences\&Engineering, and students' learning behaviors in these courses are more prosperous than those in other disciplines. The proportion of courses and student behaviors of each discipline is shown in Figure~\ref{fig:data_analysis} (a).

\subsubsection{Non-sequentiality}
\label{sec:non-sequentiality}

Considering the definition of the  \emph{Felder-Silverman Model}, we start analysis with the aspect of sequentiality. Markov property means that given the current state and all past states of a random process, the conditional probability distribution of the future state is only depends on the current state, i.e.,  conditionally independent of the historical path, which is proper to test the sequentiality. We extract each student's behavior sub-sequences of each course and integrate all that of the course to construct the transfer matrix, taking the videos involved as the discrete states.

The Markov chain of discrete sequences is often used to test the Markov property of sequences with random variables. Let $m$ denote the number of states,  $I=\{1,2,...,m\}$ denote the state space, $f_{ij}$($i,j\in I$) denote the frequency of transition from state $i$ to state $j$ in one step.  $P_{ij}=f_{ij}/\sum_{j=1}^mf_{ij}$ denotes the transition probability and $P_{\cdot j}=\sum_{j=1}^m f_{ij}/\sum_{i=1}^m\sum_{j=1}^mf_{ij}$ denotes the marginal probability. When $m$ is large, we get the chi-square test statistics $\chi^2$ of learning sequences in a course as:
\begin{equation}
    \chi^2 = 2\sum_{i=1}^m\sum_{j=1}^mf_{ij}\left|log\frac{P_{ij}}{P_{\cdot j}} \right|
\end{equation}

When the null hypothesis holds, i.e., the learning sequence is non-Markovian, $\chi^2$ will obey the chi-square distribution with the degree of freedom of $(m-1)^2$. We can obtain the chi-square value $\chi^2_{\alpha} ((m-1)^2)$ by looking up the table given the significance level $\alpha$. If $\chi^2< \chi^2_{\alpha} ((m-1)^2)$, then the null hypothesis is rejected and the learning sequence is considered to be a Markov chain.

We choose $\alpha=0.05$ as the significance level for students' learning sequences in each course and find that only 11 sampled courses meet the Markov property. Moreover, for each discipline, courses with sparse behaviors are easier to draw non-Markovian conclusions (Figure~\ref{fig:data_analysis}(a)). It demonstrates that predicting the student's behavior requires all the previous information rather than only knowing the current state, reflecting the non-sequentiality and local randomness of students' learning behaviors. 

\subsubsection{Coherence}
\label{sec:coherence}

In addition to the non-sequentiality nature revealed above, we want to explore whether internal consistency exists in students' behaviors. Consistent with course classification, we divide students into four categories according to their most learning disciplines. Figure~\ref{fig:data_analysis}(b) shows the number of students in each category, and more than half are Natural Sciences\&Engineering students.

With the assistance of course structures, we compute the proportion of four disciplines in each learning sequence. Among the 387 students, 321 have only watched videos from one discipline, revealing a global connection between behaviors. Then, we encode the subtitles using Roberta~\cite{roberta} to get the video's text representation. For a student with a sequence length of $m$, there are $m-1$ pairs of adjacent behaviors, for which we respectively calculate the inner product of these pairwise videos' representation and regard the average as the `text similarity'. When the text similarity is positive, it indicates that the adjacent videos the student has watched are similar in content.  Likewise, employing the concepts provided by MOOCCubeX, we can encode the concepts of each video using Roberta~\cite{roberta}, take their vector sum as the concept representation, and calculate the corresponding `concept similarity'.  Concept similarity is evaluated similarly to text but is more convincing since extracted knowledge filters out the colloquial expressions in videos' subtitles. We plot the number of students in each category whose concept or text similarity is positive in Figure~\ref{fig:data_analysis}(b). As we can see, most students are above the threshold, reflecting the content relevance and internal coherence of students' learning behaviors. %

\textbf{Observation 1}. \emph{Video-watching behaviors are common across students and disciplines, essential in the learning process. Students tend to fragmented learning when studying online with non-sequential behaviors. However, with the aid of course structures, text, and knowledge, we find inherent coherence in students' learning behaviors,  indicating the usefulness of these elements.} %

\section{Pre-training Implementation}
\label{sec:pretraining_model}

\begin{figure*}[t]
    \centering
    \includegraphics[width=\linewidth]{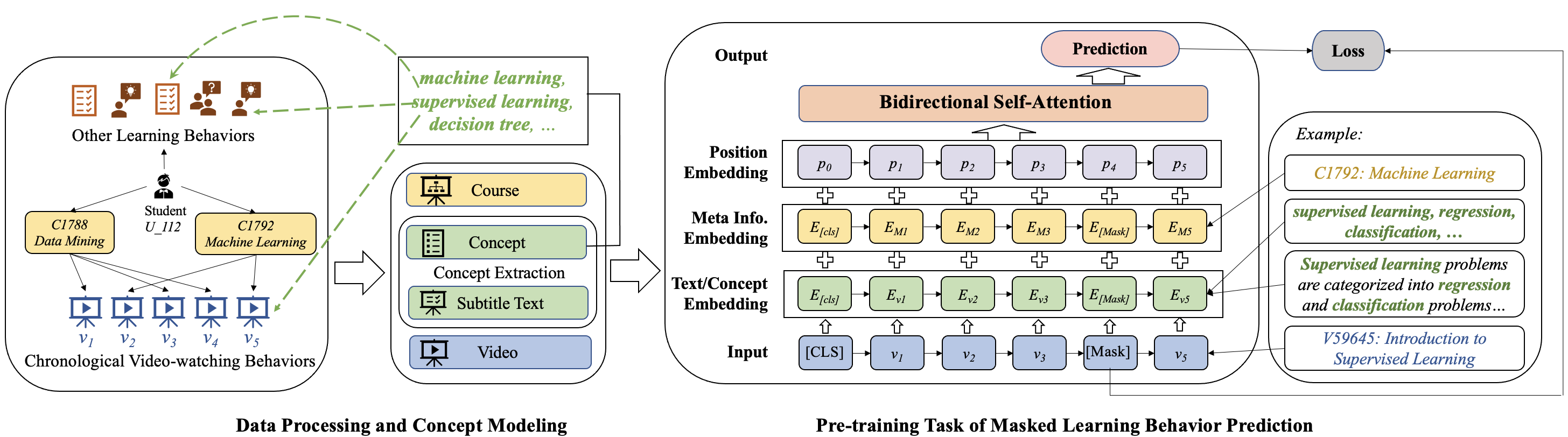}
    \caption{\centering{The overview of the pre-training framework. We collect students' learning records to form a video-watching sequence with corresponding course structures and subtitles. Then we extract high-quality concepts and link them back to textual behaviors. At last, we design a transformer-based model and conduct `Masked Learning Behavior Prediction' as the pre-training task. The rightmost is an example of the input.}}
    \label{fig:model}
\end{figure*}

Based on data observation and learning style analysis, we try to design a pre-training framework to model students' learning behaviors. First, we determine the basic model and the training objective of the pre-training stage. Then, to mitigate sparsity, we attempt to fully fuse the specific information in MOOCs to encode behavioral sequences, including course structures, text, and knowledge. Finally, we get our \textbf{P}re-training model for \textbf{A}daptive \textbf{L}earning (PAL). Figure~\ref{fig:model} shows the details of the framework. 

\subsection{Model Backbone}
\label{sec:backbone}

According to the non-sequentiality nature found in Section~\ref{sec:learning_style_analysis}, modeling students' knowledge state and preference require considering the whole learning process to leverage global information and overcome local randomness. Therefore, instead of choosing strict sequence models, we employ Transformer~\cite{attention} as the backbone, utilizing the attention mechanism to model behavioral sequences from bi-direction. Inspired by the `Masked Language Model' task proposed in BERT~\cite{devlin2019bert} we conducted a Cloze task `Masked Learning Behavior Prediction' as the pre-training objective, where a certain proportion of video-watching behaviors are randomly masked in each sequence with the specific token `[Mask]' and are predicted solely on bidirectional context for each training step. The mask proportion is preferably $0.15$, and following NLP convention, a special token `[CLS]' is added in front of each input for the sequence representation. %

The basic model we use consists of two Transformer blocks and four attention heads. Let $\mathcal{U} =\left \{u_{1},...u_{|\mathcal{U}|}\right \} $ denote the set of students, $\mathcal{V} =\left \{v_{1},...v_{|\mathcal{V}|}\right \}$ denote the set of videos, and $\mathcal{S}_{u}= \left [ v_{1}^{(u)},...v_{t}^{(u)},...v_{n_{u}}^{(u)}\right ]$ denote the chronological behavior sequence for student $u \in \mathcal{U}$, where $v_{t}^{(u)} \in \mathcal{V}$ denotes the video that $u$ has watched at time step $t$ and $n_{u}$ denotes the length of behavior sequence for $u$. The hidden vectors corresponding to `[Mask]' from the last transformer block are sent to the output layer to predict from the entire video set. Given the masked user behavior sequence $\mathcal{S}'_{u}$, the loss for every masked input is the average of the negative log-likelihood of each masked target:
\begin{equation}
    Loss = \frac{1}{|\mathcal{S}^{m}_{u}|}\sum _{v^{m}\in \mathcal{S}^{m}_{u}} - \mathtt{log}P (v^{m}=v^{\ast }|\mathcal{S}'_{u})
\end{equation}
where $\mathcal{S}^{m}_{u}$ is the set of randomly masked videos in $\mathcal{S}_{u}$, $v^{\ast}$ is the ground truth  of the masked video $v^{m}$ in $\mathcal{S}^{m}_{u}$, and $P(\cdot)$ is the probability obtained from the output layer.

\subsection{Heterogeneous Learning Elements}

Compared to the pre-training of language models, such as BERT~\cite{devlin2019bert}, which prepares a  $500G$ corpus for $30k$ tokens,  we only have about $3M$ behaviors for over $180k$ different videos. Therefore, the pre-training of behaviors faces a serious data sparsity in accord to Section~\ref{sec:data_observation}. Among the elements that previous analysis revealed helpful, course structures and text are relatively natural, while knowledge comes mainly from post-extraction. Therefore, we first conduct concept modeling and then encode learning bahaviors to our needs.

\subsubsection{Concept Modeling}
\label{sec:concept_modeling}
Concepts refer to the knowledge concepts taught in MOOCs (e.g., \emph{Binary Tree} of the course \emph{Data Structure}), which have proven essential and valuable for adaptive learning tasks~\cite{yu2019course,liu2019exploiting,wang2020neural}. As noted in section~\ref{sec:coherence}, the textual encodings of the concepts extracted in MOOCCubeX reveal content relevance of student learning behaviors. However, these results do not consider context or directly output concept representations applied to pre-training. Therefore, we re-extract high-quality concepts from learning resources and link them back to derive contextualized representations to understand learning processes better. %

\textbf{\textit{Concept Extraction}} is defined as extracting fine-grained terms or key phrases from the given learning resources~\cite{pan2017course,yu2021mooccubex}. Let $\mathcal{C} = \left \{ c_{1},...c_{|\mathcal{C}|} \right \}$ denote the concepts extracted from the given MOOC corpus $\Gamma=\{{\gamma}_{i}\}_{i=1}^{|V|}$, where ${\gamma}_{i}$ denote the subtitles of video $v_{i}$ and $\mathcal{V} =\left \{v_{1},...v_{|\mathcal{V}|}\right \}$ denote the set of videos. We employ the results from phrase mining in MOOCCubeX as ground truth to accomplish distant-supervised extraction and automatic concept labeling by fine-tuning  RoBERTa~\cite{roberta} with token classification loss, following the training scheme of Named Entity Recognition. 

Given a video $v_i$, we join its course, chapter, and title with the specific token `[SEP]' and place it before the subtitles as hints. Since the maximum input length of Roberta is $512$, we split long subtitles into segments to run prediction and select phrases with confidence larger than $0.85$ as candidates. The merged candidates are reserved as concepts for each video $v_i$, and vectors from the last hidden layer of the fine-tuned RoBERTa corresponding to the concept spans during extraction are taken as the concept representations. Compared with post Word2vec~\cite{word2vec} or text encoding after extraction, representations from the NER-fine-tuned model are more context-sensitive, highlighting the relevant background of the video. The detailed results are shown in Table~\ref{tab:concept}. 

\begin{table}[t]
\centering
\renewcommand{\arraystretch}{1.1}
\setlength{\tabcolsep}{1.2mm}
\caption{The overview of the concept extraction results. The figures in the first two columns are the average for each video, while those in the last three columns are percentages with `\%' omitted. We use Precision, Recall and F1-score to assess the extraction quality.} %
\begin{tabular}{c|c|c|c|c} 
\toprule
Avg \# of concepts & Avg \# of segments & Precision & Recall & F1-score \\ \hline
36.5 & 9.57 & 37.53 & 52.74 & 43.85 \\ \bottomrule
\end{tabular}
\label{tab:concept}
\end{table}

\textbf{\textit{Concept Linking}} is to link learning resources and behaviors to the extracted fine-grained concepts. Each video $v_i$ is linked to the concepts extracted from the subtitles $r_i$. For other behavioral objects like an exercise $e_i$ or a forum discussion $f_i$, we generate candidates and their representations from the video $v_i$ of the same learning unit and then calculate the cosine similarity between the target text and each candidate, linking the behavioral object to the most relevant $5$ concepts. %

\subsubsection{Learning Behavior Encoding}
\label{sec:behavior_encoding}

While encoding learning sequences, since the behavioral tokens of videos do not have lexical features like language models, we utilize text or extracted knowledge concepts to initialize the token embedding. Besides, inspired by the learning style analysis in Section~\ref{sec:learning_style_analysis}, we consider the characteristics of course structures and introduce positional information following previous work~\cite{sun2019bert4rec, meantime}. Thus, we can enrich the behavior encoding via heterogeneous representations to facilitate pre-training. %

\textbf{\textit{Text}}. For each video $v_{i} \in V$, we denote its subtitles as  $\gamma_i$. To represent $v_{i} \in V$ with text information, we encode $\gamma_i$ with a pre-trained language model ~\cite{roberta}. Therefore, $v_{i}$ can be represented as $PLM(\gamma_i) \in \mathbb{R}^d$ from the perspective of text.

\textbf{\textit{Knowledge Concepts}}. For each video $v_{i} \in V$, we extract and link knowledge concepts %
to construct its corresponding concept sets $C_{i} \subseteq C$. As the concept set is order-independent, we take the vector sum as the encoding result after getting the representation of each concept from the fine-tuned RoBERTa and reduce it to a low-dimension embedding through a linear layer.
Then $v_{i}$ can be represented as $\mathbf{C}_i \in \mathbb{R}^d$ from the perspective of knowledge.

Since concepts and text substitute for each other to some degree, we get the token embedding of a single video $v_i$ as: 

\begin{equation}
    \mathbf{E}_{v_i} = PLM(\gamma_i) \quad or\quad \mathbf{C}_i
\end{equation}

\textbf{\textit{Meta-information}}. 
Each video $v_{i} \in V$ belongs to a hierarchical organization structure. As introduced in Section~\ref{sec:coherence}, students' behaviors are coherent from the course level. %
Therefore, we select the corresponding course $\mathcal{M}_i$ for video $v_i$ as meta-information and encode  $\mathcal{M}_i$ into a $d$-dimensional embedding $\mathbf{E}_{M_i} \in \mathbb{R}^d$ to represent structural information.

\textbf{\textit{Position Embedding}}. To identify which portion of the input the model is dealing with, we design a learnable positional embedding matrix $P \in \mathbb{R}^{N\times d}$, where each position $i$ has a corresponding embedding $\mathbf{p}_i \in \mathbb{R}^d$.
It restricts the maximum sequence length $N$ to trade off performance against efficiency. Given a behavior sequence $\mathcal{S}_{u}= \left [ v_{1}^{(u)},...v_{t}^{(u)},...v_{n_{u}}^{(u)}\right ]$  of student $u \in \mathcal{U}$, if $n_{u}>N$, we truncate $\mathcal{S}_{u}$ to the last $N$ items $\mathcal{S}_{u}= \left [ v_{n_{u}-N+1}^{(u)},...,v_{N}^{(u)}\right ]$.

As discussed above, for a given learning sequence $\mathcal{S}_{u}$, the input representation of each behavioral token $v_{i}^{(u)}$ is constructed by summing its corresponding token embedding $\mathbf{E}^{(u)}_{v_i} \in \mathbb{R}^d$, meta-information $\mathbf{E}_{M_i} \in \mathbb{R}^d$, and position embeddings $\mathbf{p}_i \in \mathbb{R}^d$,

\begin{equation}
    \mathbf{x}_{i}^{(u)} =  \mathbf{E}^{(u)}_{v_i} + \mathbf{E}^{(u)}_{M_i} + \mathbf{p}_i
\end{equation} 

The final embedding $\mathbf{x}_{i}^{(u)}$ will be the initial hidden layer vector $\mathbf{h}_{i}^{0}$ fed into the Transformer blocks. The hidden representations $\mathbf{h}_{i}^{l} \in \mathbb{R}^d$ for each position $i$ will be stacked together into $\mathbf{H}^{l} \in \mathbb{R}^{N\times d}$ and computed iteratively at each layer $l$,  
\begin{equation}
    \mathbf{H}^{l} =  TransformerBlock(\mathbf{H}^{l-1}),\quad l \in \mathcal{L}
\end{equation}
where $\mathcal{L}$ denote the number of Transformer blocks.

To predict each masked item $v_i \in v^{m}$ base on the final output $\mathbf{h}_{i}^{L}$ of the last transformer block, we design an output layer to produce a distribution over the video set $\mathcal{V}$ for model learning:
\begin{equation}
    P(v) = Softmax(GELU(\mathbf{h}_{i}^{L}\mathbf{W}+\mathbf{b}_1)\mathbf{E}_v^T+\mathbf{b}_2)
\end{equation}
where $\mathbf{W},\mathbf{b}_1,\mathbf{b}_2$ are learnable parameters and $\mathbf{E}_v$ is the token embedding matrix described above.

\section{Experiment}
\label{sec:experiment}

According to Section~\ref{sec:data_analysis}, we find that students in MOOCs tend to learn in fragments non-sequentially and that the elements of course structures, text, and knowledge are helpful and inherently coherent to students' learning behaviors. A systematic analysis of pre-training for adaptive learning through experimentation is warranted to understand the specific roles of different elements and how multiple types of information work together. In order to answer the three critical questions proposed in the opening section, we select tasks targeting each component of adaptive learning respectively and further discuss in light of the results.

\subsection{Overview}

\sequence~aims to analyze and predict the students' behaviors, highly relevant to the pre-training task of `Masked Learning Behavior Prediction'. We employ \emph{Learning Recommendation} as the main experiment to illustrate the performance and efficiency of pre-training.
\content~aims to model the learning resources. We conduct a \emph{Learning Resource Evaluation} task for the \emph{expert module} to explore utilizing the pre-training model as an encoder. 
\assessment~focuses on modeling the students. \emph{Knowledge Tracing} is a typical high-demanding low-resource task and models the student's state of knowledge based on historical learning records at the micro level, while \emph{Dropout Prediction} is instead a macro inference of whether the student will complete the course. We choose these two tasks to evaluate the transferability of the pre-training method. 
To accommodate each task, a series of datasets based on MOOC2.7M are necessities. In summary, the adaptive learning tasks selected for experiments with their corresponding datasets are shown in Table~\ref{tab:task}. %

\textbf{Running Environment.} The experiments in this section are conducted on a single Linux server with an Intel(R) Xeon(R) CPU E5-2669 v4 @ 2.20GHz, 256G RAM, and $8$ NVIDIA GeForce TITAN X (Pascal). The codes of our proposed models are implemented with Pytorch 1.7.0 in Python 3.7.   

\begin{table*}[t]
\centering
\renewcommand{\arraystretch}{1.1}
\setlength{\tabcolsep}{1.2mm}
\caption{The adaptive learning tasks chosen for exploring the effectiveness of pre-training. `/' means the information is not directly modeled in the task.} %
\begin{tabular}{c|c|c|c|c|c|c} 
\toprule
Direction  &  Task  & Dataset & \# of students & \# of videos & \# of courses & \# of video-watching behaviors\\ \hline
\sequence  & \emph{Learning Recommendation} & MOOC2.7M & 117,992 & 181,951 & 4,124 & 2,680,833\\ \hline
\content & \emph{Learning Resource Evaluation} & MOOC-LRE &  / & 1,422 & 8,269 & / \\ \hline %
\multirow{2}{*}{\begin{tabular}[c]{@{}c@{}}\assessment \end{tabular}} & \emph{Knowledge Tracing}    & MOOC-KT & 2,093 & 857 & 12 & 217,960\\ \cline{2-7}
& \emph{Dropout Prediction}  & MOOC-DP & 117,992 & / & 4,124 & / \\  \bottomrule
\end{tabular}
\label{tab:task}
\end{table*}

\subsection{Performance \& Efficiency}

\emph{Learning Recommendation} aims to recommend the following videos for students to study based on their historical learning records. As defined in Section~\ref{sec:backbone}, given the interaction history $\mathcal{S}_{u}$, it predicts the video that user $u$ will learn at time step $n_{u}+1$, which can be formalized as modeling the probability over all possible videos for user $u$ at time step $n_{u}+1$.

\subsubsection{Experimental Settings and Baselines}

To evaluate the \emph{Learning Recommendation} models, we adopt the leave-one-out evaluation, i.e., the next video recommendation task, which has been widely used in \cite{he2017neural,kang2018self,tang2018personalized,sun2019bert4rec}. For the behavior sequence of each user, we treat the last video as the test set, the penultimate video as the validation set, and use the remaining sequence for training. Following the existing works~\cite{he2017neural,tang2018personalized,sun2019bert4rec}, we evaluate our model with Popular Negative Sampling, where each video in the test set is ranked with the $100$ most popular negative videos that the student has not watched. We typically employ Normalized Discounted Cumulative Gain (NDCG) and Recall as the evaluation metrics and report the top-k result (k = $1$, $5$, $10$). The higher values for these metrics, the better model performance is. Note that since we have only one ground-truth video for each user, NDCG@$1$ is equivalent to Recall@$1$.

Our pre-training method is compared with five representative educational recommendation baselines:

$\bullet$ \textbf{POP}, a simple baseline which ranks the videos according to their popularity of interactions;

$\bullet$ \textbf{KSS}~\cite{liu2019exploiting}, ranking the videos according to the course structures, which is another simple baseline for learning decision;

$\bullet$ \textbf{GRU4Rec}~\cite{hidasi2015session}, a widely-applied session-based recommendation model with GRU architecture;

$\bullet$ \textbf{CASER}~\cite{tang2018personalized}, employing CNN to model high-order MCs in both horizontal and vertical way for sequential recommendation; %

$\bullet$ \textbf{BERT4Rec}~\cite{sun2019bert4rec}, a classical BERT-like pre-training method for recommendation based on bidirectional Transformer.

\begin{table}[t]
\centering
\setlength{\abovecaptionskip}{0.2cm}
\setlength{\belowcaptionskip}{0.cm}
\caption{Results of \emph{Learning Recommendation} with Popular Negative Sampling. All figures in the table are percentages with `\%' omitted.} %
\label{experimental_result}
\begin{tabular}{c|c|cc|cc}
\toprule
\multirow{2}{*}{Model} & Metric@1 & \multicolumn{2}{c|}{Metric@5} & \multicolumn{2}{c}{Metric@10} \\ 
\cline{2-6}%
& NDCG     & NDCG         & Recall         & NDCG          & Recall         \\ 
\hline%
POP     &  \ \ 0.02  &  \ \ 0.05 &  \ \ 0.08   &  \ \ 0.07  &  \ \ 0.14                \\ %
KSS  &  \ \ 0.98   &  \ \ 2.90  &  \ \ 4.94  &  \ \ 4.50  &  \ \ 9.94                \\ %
GRU4REC &  45.14 &   53.47    &  61.55     & 61.79  &    65.81                  \\ %
CASER  & 33.57 &   46.51   &  48.87     &  59.46  &    66.84               \\ %
BERT4Rec  & 60.34 & 65.13 & 69.34 & 66.12 & 72.40                \\ %
\hline%
\model & \textbf{70.94} & \textbf{74.73} & \textbf{77.96} & \textbf{75.43} & \textbf{80.11}   \\ \bottomrule
\end{tabular}
\end{table}

\subsubsection{Results} 

From the results in Table \ref{experimental_result}, we can find that:

(1) The simple baselines POP and KSS perform much worse than neural methods since they do not use historical learning records to model students' personalized preferences. 

(2) Methods based on pre-training models, i.e., BERT4Rec and \model, outperform traditional sequential models, i.e., GRU4Rec and CASER. As student behaviors in actual MOOC scenarios have non-strict sequential patterns, thus compared to the sequential models that only learn from supervised signals, pretraining-based methods can mine in-depth connections between student behaviors from large-scale unlabeled data and are more suitable for solving behavior prediction tasks.  %

(3) \model~significantly outperforms BERT4Rec, even with a similar pre-training task, further demonstrating the effectiveness of heterogeneous learning elements. While BERT4Rec only considers the sequence of student interactions with specific videos,  \model~incorporates more complex correlations and appropriately utilizes the various learning elements in the pre-training process, thus achieving notable performance improvements.

\subsubsection{Ablation Study}
\label{sec:ablation_study}

\begin{table}[t]
\caption{Results of Ablation Study. `Conc.', `Text', `Meta.' correspond to \model~with concepts, text, meta-information. Concepts and text substitute for each other, so they can not be used simultaneously.}%
\label{ablation}
\resizebox{\linewidth}{!}{
\begin{tabular}{c|c|cc|cc}
\toprule
\multirow{2}{*}{Setting} & Metric@1 & \multicolumn{2}{c|}{Metric@5} & \multicolumn{2}{c}{Metric@10} \\ \cline{2-6} 
 & NDCG     & NDCG         & Recall         & NDCG          & Recall         \\ \hline
+Conc.   &  +4.12 &  +3.81  &  +3.45  &  +3.64  &  +2.92                \\ \hline
+Text    &  +6.24 & +5.92 & +5.54 & +5.74 & +4.96                \\ \hline
+Meta.   &  +9.59 &  +9.01  &  +8.42  &  +8.73  & +7.52                \\ \hline
+Conc.\&Meta. & +10.47 & +9.48 & +8.56 & +9.17 & +7.60                \\ \hline
+Text\&Meta. &  \textbf{+10.60} & \textbf{+9.60} & \textbf{+8.62} & \textbf{+9.31} & \textbf{+7.71}  \\ \bottomrule
\end{tabular}}
\end{table}

To further verify the effectiveness and explore the role of each element considered, we conduct a thorough ablation study. Affiliated modules are added one by one to the pre-training task of `Masked Learning Behavior Prediction' to confirm the related improvements brought by each module. By comparing the results of different settings in Table~\ref{ablation}, we have the following observations: 

(1) Each of the affiliated module and additional learning element we add for \model~positively affects the performance. 

(2) Meta-information from course structures boosts more pronounced than concepts and text by 5.47\% and 3.35\% for NDCG@$1$. As observed in Section~\ref{sec:coherence}, most students only watch videos from a single discipline. Better organized and structured meta-information facilitates the model to overcome the local randomness of non-sequential learning behaviors. %

(3) Using text or concepts with meta-information yields similar results, i.e., a difference of 0.13\%, despite a difference of 2.12\% when the two are used alone. It indicates that the extracted knowledge concepts cover the core content of videos, although the non-core content in the text also reflects a consistent style. With meta-information increasing knowledge centralization, concepts can bring clearer interpretability to adaptive learning tasks~\cite{yu2019course} without sacrificing the model's effectiveness.

\subsubsection{Training Efficiency Analysis}

\begin{figure*}[t]
    \centering
    \includegraphics[width=0.7\textwidth]{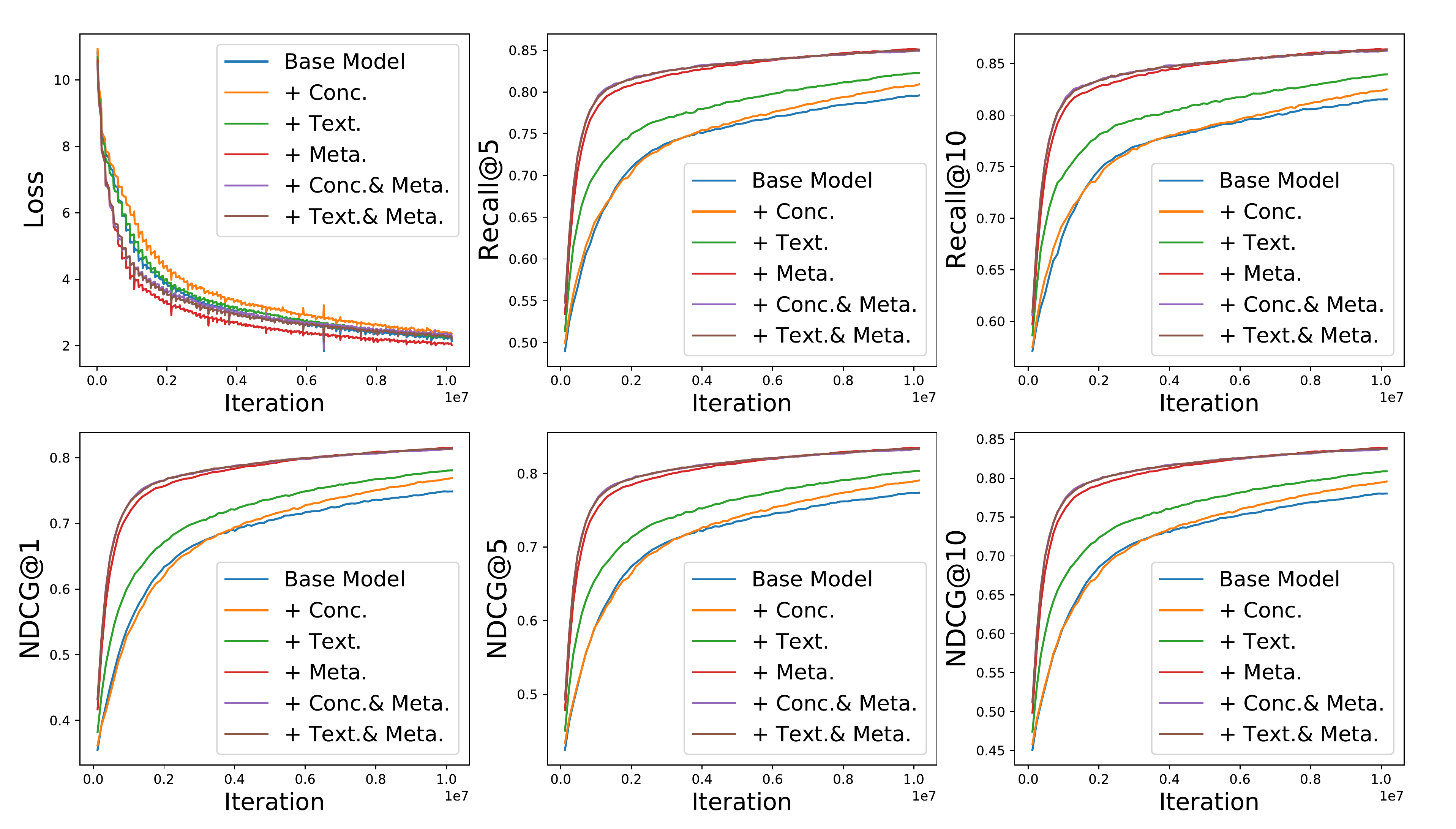}
    \caption{The loss and model performance curves of \model~ during training in \emph{Learning Recommendation} under different ablation settings.}
    \label{fig:train}
\end{figure*}

Training efficiency is a common concern with pre-training methods, as massive data and parameters are computing-consuming. Newly added modules are expected to improve performance as well as training efficiency. Figure \ref{fig:train} shows the loss and performance curves during training, from which we can find that: 

(1) The convergence is significantly faster when using meta-information.  As the amount of courses is much less than that of videos, adding course structures increases the density of knowledge and helps mitigates data sparsity.

(2) The `only concepts' setting converges slightly slower than the base model, possibly because the condensed content represented by concepts amplifies the differences between videos. Therefore, it is more feasible and efficient to adopt text than concepts in pre-training while directly applying background knowledge to downstream tasks. %

\textbf{Observation 2}. \emph{\textbf{Meta-information enhances knowledge concentration and reduces local randomness of learning behaviors, while text and concepts help model the consistency of learning content}. In line with pedagogy views, learning behaviors include multiple influences in addition to resource attributes and student profiles. Experiments confirm that heterogeneous learning elements included in pre-training can significantly promote efficacy.}

\subsection{Facilitating Downstream Tasks}

\subsubsection{Encoding Approaches} %

Traditionally, instructional materials are primarily evaluated manually by experts~\cite{tyler1967changing}. However, with the upsurge of \content, automated evaluation of learning resources has become a rising topic in recent years. \emph{Learning Resource Evaluation} concerns modeling the correlation between the representation of resources and the corresponding indicators. Let $\mathcal{E}$ denote a set of learning resources and $y_{i}$ denote an attribute indicator of resource $e_i \in \mathcal{E}$. If $y_i$  is a continuous variable, the prediction task will be transformed into an $N$-classification problem to mitigate noise. $y_{i} = j (j \in \{1,..,N\})$ indicates that the resource $e_i$ belongs to class $j$. We perform the \emph{Learning Resource Evaluation} task at both the course-level and the video-level.

\noindent $\bullet$ \textbf{Experimental Settings} %

\noindent \emph{Course Completion Rate} is one of the critical indicators to evaluate the quality of a course, widely accepted for learning analytics. After filtering out courses with a low number of students, we calculate the average completion rate of each course, i.e., the quotient of the number of students who watched all videos divided by those who registered. For the $1,422$ courses eventually retained, we find that the logarithm of course completion rate conforms to normal distribution. Therefore, we group the logarithm course completion rate into four buckets according to the top 25\%, 25\%-50\%,  50\%-75\%, and the remaining 25\%, to reflect the long-term appeal of the course content to students.

\emph{Video Comment Rate} is an indicator of the quality of a video. Students can engage in intellectual discussions in forums attached to the videos in MOOCs. The comment rate, i.e., the number of comments per student watching the video, is considered to strongly correlate with the attractiveness of the video content~\cite{chen2014investigating}. Therefore, we filter out videos with few play counts or no comments, leaving $8,269$ videos, and group them regarding the logarithm video comment rate into four buckets as above. 

For each course, we pack the course videos into a sequence and take the final hidden vector of the special `[CLS]' token at the front as the course encoding, while for each video,  we directly extract the corresponding token embedding from the trained \model~as the video encoding. For the experiments, we use XGBoost as the classification model and leverage bayesian optimization to search for the best hyperparameters. All datasets are split into train/evaluation/test by 8:1:1. For the above two sub-tasks, we choose Accuracy, Precision, Recall, and macro F1-score as evaluation metrics for the four-class classification and select the best model based on the F1-score on the evaluation set.

\begin{table}[t]
\setlength{\abovecaptionskip}{0.2cm}
\setlength{\belowcaptionskip}{0.cm}
\setlength{\tabcolsep}{1.2mm}
\caption{Results of the \emph{Learning Resource Evaluation} task. \\The figures in the table are percentage numbers with `\%' omitted.}
\label{tab:regression}
\centering
\begin{tabular}{c|cccc}
\toprule
Setting & Accuracy & Precision &  Recall & macro-F1 \\\midrule
\emph{Course Completion Rate} & 40.18 & 40.33 & 40.84 & 40.42  \\
\emph{Video Comment Rate} & 51.63 & 48.11 & 49.37 & 48.05   \\
\bottomrule
\end{tabular}
\end{table}

\begin{figure}[t]
    \centering
    \includegraphics[width=0.95\linewidth]{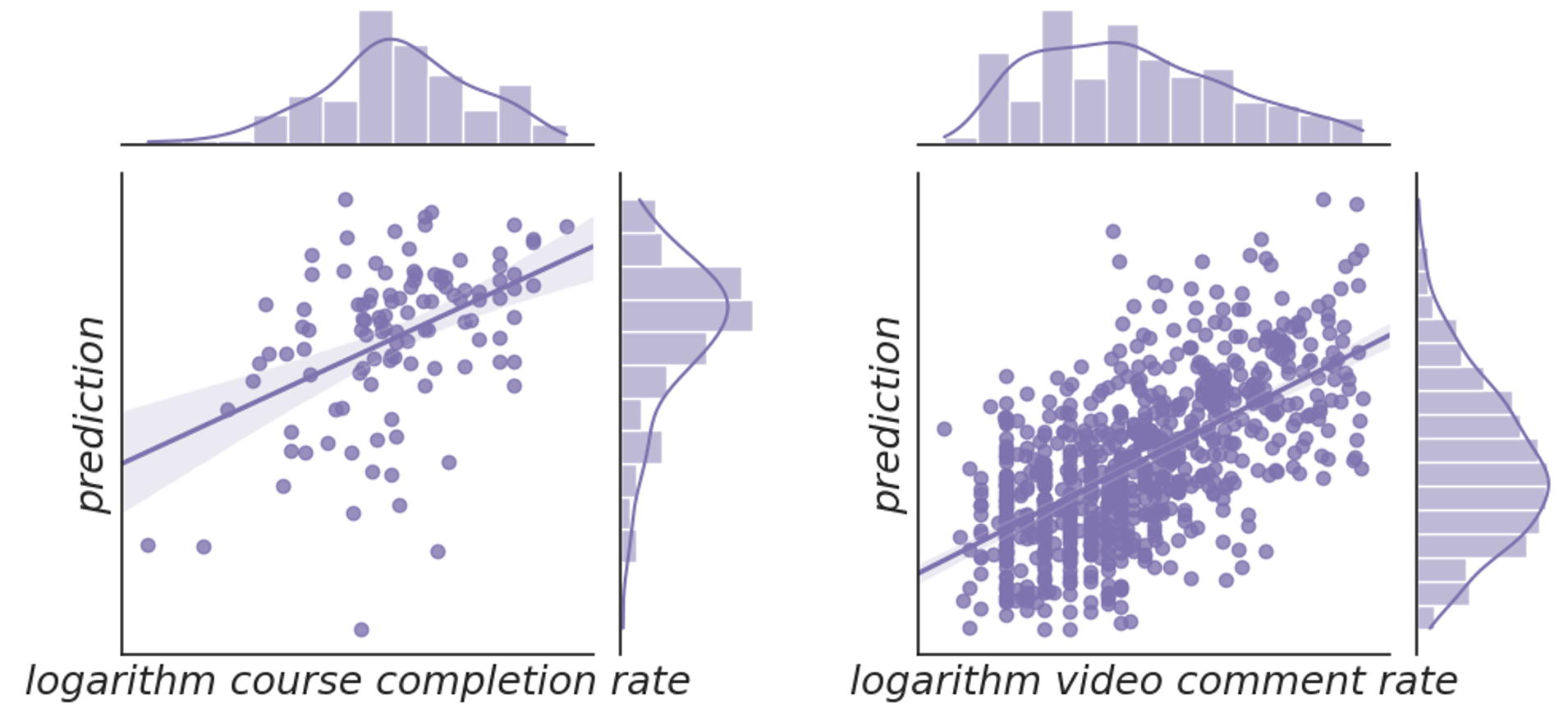}
    \caption{The distributions of the ground truth and prediction for \emph{course completion rate}(left) and \emph{video comment rate}(right)  show consistency.}
    \label{fig:regression}
\end{figure}

\noindent $\bullet$ \textbf{Experimental Results} %

\noindent From Table~\ref{tab:regression} and Figure~\ref{fig:regression}, we can see that: 

(1) For a four-classification problem, an F1-score over 0.4 indicates that the prediction is pretty good. Whether extracting video encoding directly from the embedding layer or packing the videos of a course into a sequence to obtain the course encoding, these dense representations show great semantic expression. 

(2) The prediction and ground truth distributions reflect good consistency for course completion rate and video comment rate. Meanwhile, the content evaluation of videos performs better than courses, possibly because the course dataset is small, limiting the performance of the classification model. Looking into the wrong examples, most are near the separation surface. It indicates that the classification result can be better, since the pre-set classification threshold may inhibit the current result.

\begin{figure}[t]
    \centering
    \includegraphics[width=0.85\linewidth]{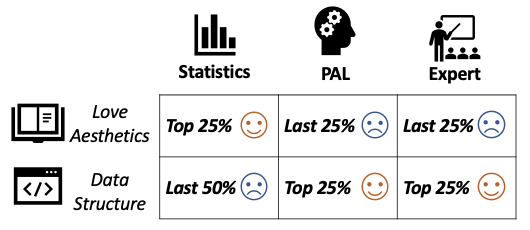}
    \caption{\centering{An example of \model~evaluating better than statistical indicators.}}
    \label{fig:app_example}
\end{figure}

\noindent $\bullet$ \textbf{Qualitative Analysis} 

\noindent In addition to the quantitative results, we further conduct a case study. To our surprise, while statistical methods based on pedagogical theory may fail when evaluating resources, \model~can evaluate closer to expert annotation. For example, the predicted comment rate of video \emph{Love Aesthetics} with inactive watching behaviors belongs to the last 25\% by \model, while the statistical rate is at the top 25\%. According to students' feedback, the video is ordinary, but it has additional requirements that participation in forum discussions is part of the scoring method. Therefore, the result of \model~coincides more with the actual quality of the video. %

Another example is that course \emph{Data Structure} is an award-winning excellent course, of which the predicted course completion rate is top 25\%, while the rate in statistics is 50\%-75\%. According to the data analysis, most students on XuetangX have registered for this popular course. However, students with diverse backgrounds and knowledge levels do not have the interest and ability to study continuously and drop out midway. It shows that statistical indicators tend to fail in case of noise. The representation obtained from \model~performs more steadily and provides a robust method for \emph{Learning Resource Evaluation}.

\textbf{Observation 3}. \emph{\textbf{The pre-trained adaptive learning model preserve an excellent ability in encoding the learners as well as teaching materials}, regardless of whether the encoding approach is the embedding layer or sequence encoding. Pre-training comprehensively models various learning elements, resulting in a stable and robust representation. While pedagogical statistical evaluation may fail under certain circumstances, the pre-training method can evaluate closer to the actual quality of the resource.}  %

\subsubsection{Low-resource Scenarios}

\emph{Knowledge Tracing} concerns three main entities: student, question and concept. As defined before, $\mathcal{U}$ denotes the set of students and $\mathcal{C}$ denotes the set of concepts. Besides, let $\mathcal{Q}$ denote the set of questions and $y_{ij} \in \{0, 1\}$ denote the result of student $u_i \in \mathcal{U}$ answering question $q_j \in \mathcal{Q}$, where $y_{ij} = 1$ indicates a correct answer and $y_{ij} = 0$ indicates a wrong answer. 

\noindent $\bullet$ \textbf{Experimental Settings and Baselines}

\noindent We select $12$ computer science courses from MOOC2.7M and create a specific dataset MOOC-KT for the \emph{Knowledge Tracing} task. Apart from the statistics presented in Table~\ref{tab:task}, the dataset involves $519$ questions, $101$ concepts, and $116,790$ question-doing records. We pick several representative baselines for comparison.

$\bullet$ \textbf{MIRT}~\cite{reckase2009multidimensional}, a method of education data mining that extends the latent trait value of each student in IRT~\cite{drasgow1990item} to a multi-dimension knowledge proficiency vector with the Q-matrix. 

$\bullet$ \textbf{DKT}~\cite{piech2015deep}, a widely-used knowledge tracing method employing RNNs to learn the complex interactions from exercise records.

$\bullet$  \textbf{PDKT}~\cite{chen2018prerequisite}, a DKT-based method employing the prerequisite relation of concepts to optimize modeling for better performance.%

$\bullet$ \textbf{NCD}~\cite{wang2020neural}, a cognitive modeling method that incorporates neural networks to learn complex exercise interactions.%

Most baseline methods do not consider student learning behaviors, except that the current state-of-the-art NCD has a module on student profiling through which we can engage the pre-trained model in downstream training. We feed the videos that the student has learned into \model, encode the sequence, and take the hidden vector of `[CLS]' token as the initialized representation of the student, further fine-tuned with NCD. Following the existing works~\cite{piech2015deep, chen2018prerequisite, wang2020neural}, we choose Accuracy (ACC), Root-mean-square deviation (RMSE), Area under the ROC Curve (AUC) and macro F1-score (F1) as evaluation metrics. Higher values of ACC, AUC, and F1-score imply better model results, while the opposite is true for RMSE. Ultimately, the best model is selected based on the F1-score on the evaluation set. %

\begin{table}[t]
\setlength{\abovecaptionskip}{0.2cm}
\setlength{\belowcaptionskip}{0.cm}
\setlength{\tabcolsep}{1.2mm}
\caption{Results of the \emph{Knowledge Tracing} task compared with baselines.}
\label{tab:lkt}
\centering
\begin{tabular}{l|cccc}
\toprule
Methods     &  ACC &  RMSE      &  AUC   & F1 \\\midrule
MIRT     &   63.11 & 46.72 & 60.31             &   50.12     \\
DKT        & 77.02 & 39.91 & 74.95             & 64.24    \\
PDKT       & 77.23 & 40.01 & 74.21             & 64.37  \\ \midrule
NCD   & 81.48       & 38.68      & 77.57     & 68.42  \\ 
NCD$_{\model}$ & \textbf{83.41} & \textbf{34.86}   & \textbf{79.86} & \textbf{69.17}  \\
\bottomrule
\end{tabular}
\end{table}

\noindent $\bullet$ \textbf{Experimental Results} %

\noindent From the results shown in Table \ref{tab:lkt}, we can see that:  

(1) NCD does outperform the other baseline approaches, which is justified because it explicitly maintains a profile of the student in addition to modeling the student's exercise records.

(2) Employing \model~to encode learning behaviors as the student profiling module for NCD can additionally improve the results. It demonstrates that though \model~primarily models students' video-watching behaviors rather than the exercising behaviors used directly in \emph{Knowledge Tracing}, the two types of behaviors are closely correlated. It may be because students' knowledge mastery can be inferred from the content and knowledge concepts they have learned, which also indicates that the modeling ability of \model~is powerful and transferable.

\begin{table}[!t]
\setlength{\tabcolsep}{1.2mm}
\caption{Performance of \emph{Knowledge Tracing} at different low-resource settings. Figures in ACC, AUC, and F1 are percentages with `\%' omitted.}
\label{tab:lkt}
\centering
\begin{tabular}{c|l|cccc} \toprule %
 Setting  & Methods   & ACC & RMSE & AUC & F1 \\ \midrule
\multirow{3}{*}{100\%} 
 &   NCD   & 81.48       & 0.3868      & 77.57     & 68.42    \\ 
 & NCD$_{\model}$   & \textbf{83.41} & \textbf{0.3486}   & \textbf{79.86} & \textbf{69.17}    \\ \cline{2-6} %
  & \emph{Gain} & (+1.93) & (-0.0382) & (+2.29) & (+0.75) \\ 
 \midrule
\multirow{3}{*}{30\%}   
 &   NCD                     & 82.37       & 0.3626    & 78.92     & \textbf{68.88}    \\
 & NCD$_{\model}$    &  \textbf{84.48}   & \textbf{0.3448}      & \textbf{79.86}     & 68.25    \\\cline{2-6}
  & \emph{Gain}  & (+2.11) & (-0.0178) & (+0.94) & (-0.63)  \\ \midrule %
\multirow{3}{*}{10\%}  
 &  NCD   & 78.50       & 0.4292     & 69.42     & 61.97 \\  
 & NCD$_{\model}$    &  \textbf{82.03} & \textbf{0.3889}   & \textbf{74.58} & \textbf{65.40}    \\\cline{2-6}
  & \emph{Gain} & (+3.95) & (-0.0403) & (+5.16) & (+3.43)   \\ \bottomrule
\end{tabular}
\end{table}

\noindent $\bullet$ \textbf{Low-resource Study}

\noindent \emph{Knowledge Tracing} is a low-resource task because of its high demand on the dataset, which requires adequate records of student learning behaviors, especially exercises, and expertly annotated knowledge concepts for each question. Therefore, we investigate the relative improvement brought by \model~in different low-resource scenarios. Table \ref{tab:lkt} shows the performance of NCD and NCD$_{\model}$ with only $10\%$, $30\%$, and $100\%$ of the training data retained. %

As can be seen, the performance of both methods drops with the decrease of the training data. However, the relative gain in Accuracy of NCD$_{\model}$ over NCD increases in each setting, i.e., 1.93\%, 2.11\%, 3.95\%. Although such a phenomenon is not surprising, it is crucial since low-resource scenarios are widespread in adaptive learning. In the extreme case of only 10\% data retained, where the student may only have one record,  NCD$_{\model}$ performs steadily and achieves superior results on all metrics, suggesting that PAL allows us to make accurate interventions in advance to benefit student learning~\cite{shah2020year}. %

\textbf{Observation 4}. \emph{\textbf{Student learning behaviors can help infer their mastery of knowledge}. The Knowledge Tracing task demonstrates that though the video-watching behaviors seem not to match the question-doing behaviors directly, they imply students' preferences and knowledge states from different perspectives that can complementarily benefit from joint modeling and achieve steady performance in low-resource scenarios.}

\subsubsection{Long-term traits of students} 
\emph{Dropout Prediction} pays attention to long-term traits rather than behaviors on specific videos or questions. Let $\mathcal{U}$ denote the set of students as before, $\mathcal{M}$ denote the set of courses, and the pair $(u_i, m_j)$ denote that student $u_i \in \mathcal{U}$ enrolls in the course $m_j \in \mathcal{M}$. Given student $u_i$'s learning activities for course $m_j$ over the historical period, $y_{ij} \in \{0, 1\}$ denotes whether $u_i$ will dropout from  $m_j$ in the prediction period, where $y_{ij} = 1$ indicates a withdrawal behaviour and $y_{ij} = 0$ the opposite.

\noindent $\bullet$ \textbf{Experimental Settings}

\noindent  We select a representative method~\cite{feng2019understanding} for \emph{Dropout Prediction} in MOOCs as the baseline, which considers diverse information such as learning activities, learning intervals, and student social networks to construct several statistical features for training the classifier. Similar to the \emph{Knowledge Tracing} task, we input the videos that a student has learned into \model~and take the sequence encoding as the additional features of the student. The validity of the features are tested with two widely-used classification models: Logistic Regression (LR) and multi-layer Deep Neural Network (DNN), with AUC and average precision (AP) as metrics.

\begin{table}[t]
\setlength{\abovecaptionskip}{0.2cm}
\setlength{\belowcaptionskip}{0.cm}
\setlength{\tabcolsep}{1.2mm}
\caption{Result of \emph{Dropout Prediction}.  `only \model' means only employing the features provided by \model. `only Origin' means only employing the features proposed in the baseline. `Ensemble' means the ensemble results. $\Delta$ means the improvement over `only Origin' after ensemble.} %
\label{tab:dropout}
\centering
\begin{tabular}{c|c|c|c|c|c}
\toprule
Methods  & Setting      &  AUC   &  $\Delta$ &  AP   &  $\Delta$ \\ \midrule
\multirow{3}{*}{LR}       & only \model & 67.3 & -    & 85.6   & -\\
 & only Origin  & 72.6 & -    & 86.5  & -\\ 
         & Ensemble & \textbf{76.6} & \textbf{+4.0} & \textbf{89.8}     &  \textbf{+3.3}\\\midrule
\multirow{3}{*}{DNN}      & only \model  & 67.4 & -    & 86.0   & -\\ 
 & only Origin & 73.1 & -    & 87.0   & -\\
         & Ensemble &76.5 & \textbf{+3.4}  &  89.5 & \textbf{+2.5} \\
\bottomrule
\end{tabular}
\end{table}

\noindent $\bullet$ \textbf{Result Analysis}

\noindent From the results in Table~\ref{tab:dropout}, we focus on two phenomena: 

(1) Employing features provided by `only \model' as input is competitive in AP with `only Origin', whose features are proposed based on pedagogical theory. It suggests that \model~can capture students' long-term interests and learning patterns by modeling fine-grained sequential behaviors to speculate whether the student will persist in the course over time. %

(2) Compared to the baseline, which focuses on direct statistical information, time intervals between behaviors, and the influence of friends, PAL implicitly models the student's learning content and explores the correlation between his/her performance in different enrolled courses. Both methods complement each other, so using PAL's output as additional features can lead to better ensemble results with different classifiers. It indicates pre-training can mine deeper relations beyond human annotation and effectively supplement current statistical methods.

\textbf{Observation 5}. \emph{\textbf{Modeling fine-grained sequential behaviors on specific videos can capture students' long-term interests and macro learning patterns}. The Drop Prediction task reveals that by implicitly modeling the learning content and student activity on different learning resources during pre-training, the pre-trained model can mine deeper long-term traits of the student and compensate for existing statistical methods.}

\subsection{Practical Adaptive Application} 

\begin{figure*}[t]
    \centering
    \includegraphics[width=0.95\linewidth]{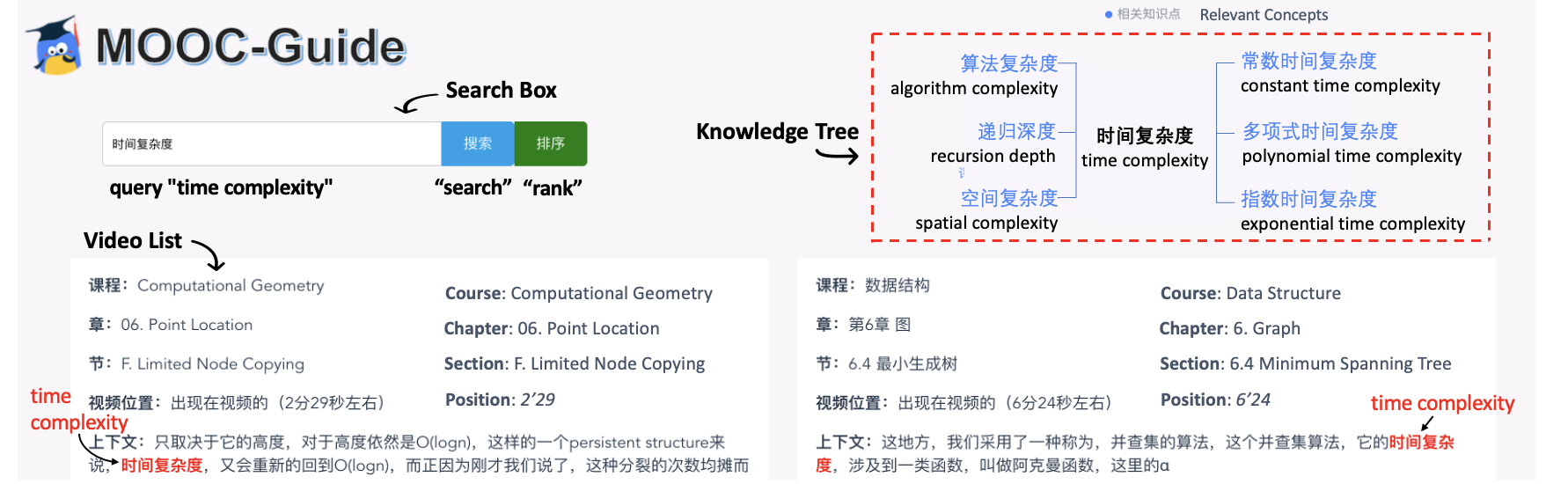}
    \caption{\centering{A screenshot of \system~deployed at XuetangX. When a student queries in the search box, the knowledge tree will appear in the upper right. Related videos will be listed below with the concept's position, ranking by knowledge relevance or learning behavior modeling.}}
    \label{fig:moocguide}
\end{figure*}

Building general adaptive systems is the ultimate goal of adaptive learning, which aims to plan learning paths for individual learners. We employed \model~to build a resource retrieval and recommendation tool, \system~on XuetangX, as a simplified systematic application to validate the practicality of pre-training. Figure~\ref{fig:moocguide} is a screenshot. When a student queries in the search box, the corresponding knowledge point and relevant concepts will appear in the upper right, and related videos will be listed below, sorted by knowledge relevance, with the location of the concept indicated. When the student clicks on the `rank' button, \system~will recommend videos based on his/her historical learning behaviors. For example, given query `time complexity' in Figure~\ref{fig:app_example}, for a math student, \system~will recommend high-level math-related videos of \emph{Computational Geometry}, while for a CS student, it will recommend videos introducing in-depth algorithms. %

\begin{figure}[t]
    \centering
    \includegraphics[width=\linewidth]{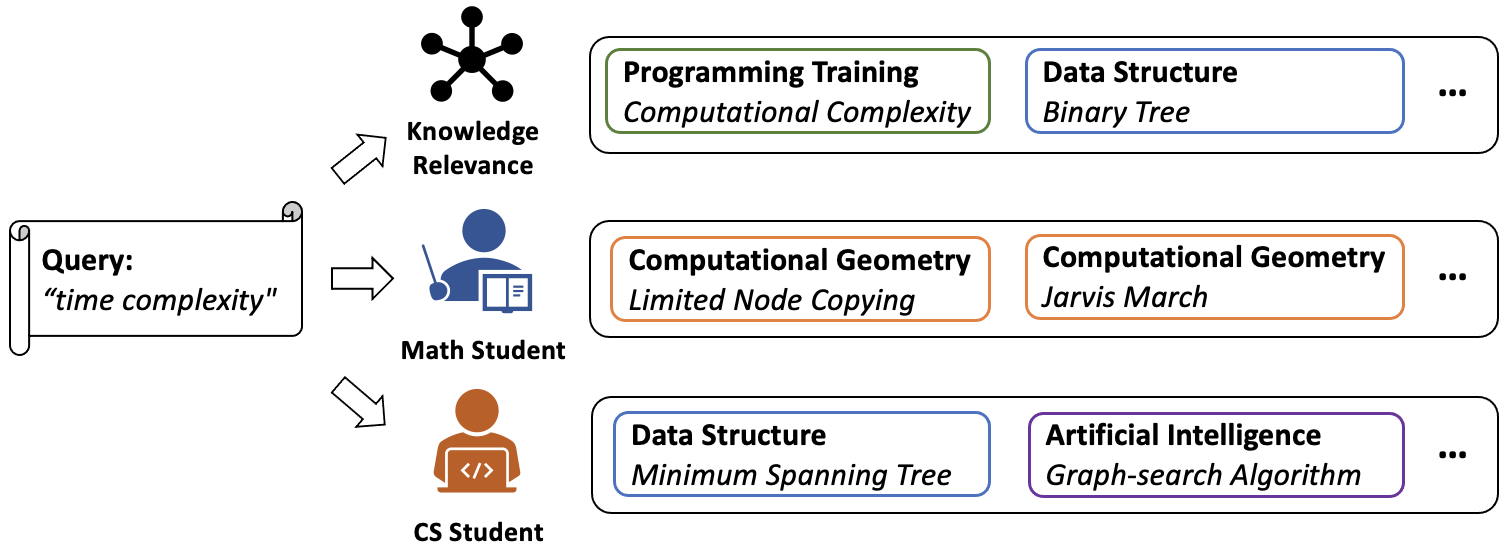}
    \caption{\centering{An example of \system, given query  `time complexity'. The displayed video list is different, ranking by knowledge relevance or personalized recommendation for a math or  CS student. Each box is a video with the course name in \textbf{bold} and the video name in \emph{italics}.}}
    \label{fig:app_example}
\end{figure}

Furthermore, we conduct an A/B test on four courses (i.e., \emph{Financial Analysis and Decision Making}, \emph{Introduction to Psychology}, \emph{C++ Programming}, and \emph{Data Structure}) to test the actual effectiveness of \system. We randomly split students into a treatment group and a control group. After two weeks of data collection, we examined the video-watching and question-doing behaviors of the different groups. As shown in Table~\ref{tab:abtest}, \system~is effective in increasing students' video engagement and correct response rates. As a result, \system~has subsequently been deployed on more than $200$ courses on XuetangX and currently serves over $20,000$ students. 

\textbf{Observation 6}. \emph{\textbf{The systematic application of adaptive learning can bring personalized intervention to students to improve video engagement and correct answer rates}. It indicates that the pre-trained model can be deployed in real-world online learning environments to provide convenient resource retrieval and recommendation and that adaptive learning can better meet students' needs and stimulate their initiative and enthusiasm, thus promoting learning outcomes.}

\begin{table}[!t]
\setlength{\tabcolsep}{8pt}
\caption{Results of intervention by A/B test. WVT: average time (s) of video watching; CAR: average rate of correct answers. $\ast $: $p$ value $\leq 0.1$.}
\label{tab:abtest}
\centering
\begin{tabular}{@{}c|c|c@{}}
\toprule
Activity & WVT & CAR \\ \midrule
No Intervention      & 4,926.48    &    0.29 \\ 
With Intervention      & 4,998.03   & 0.40*  \\ \bottomrule
\end{tabular}
\label{dkt}
\end{table}

\section{Discussion and Impact Analysis}

Based on data analysis and systematic experiments, we gained several observations and reflections about the critical questions of pre-training in adaptive learning:

$\bullet$ Learning behaviors include multiple influential elements, where course structures, text, and knowledge are helpful and inherently coherent to students’ non-sequential learning behaviors. Meta-information of courses increases knowledge concentration, while text and concepts can enhance the consistency of learning content. Proper and adequate use of the extensive information can promote effective modeling of learning behaviors.

$\bullet$ Indirectly relevant information included in the pre-training foundation can be shared across a series of downstream tasks to promote effectiveness, and unified pre-training can better integrate diverse components into a general adaptive system. For example, \emph{Knowledge Tracing} demonstrates that video-watching behaviors can help infer student knowledge mastery; and pre-training can capture long-term learning patterns by implicitly modeling learning content and performance across courses in \emph{Drop Prediction}. Pre-trained models as encoders also show great semantic expression, regardless of whether the encoding is from the embedding layer, sequence encoding, or post-fine-tuning.

$\bullet$ In alignment with pedagogy views, adaptive learning can meet student needs and facilitate learning outcomes through resource retrieval and recommendation. The simplified systematic application indicates that pre-trained models can adapt to the real-world MOOC platform. On the other hand, \emph{Learning Resource Evaluation} suggests that pre-training evaluations are sometimes even closer to expert annotation than pedagogical statistics, which prompts us to reflect on reforms in educational evaluation. %

\section{Conclusion}
\label{sec:conclusion}
As far as we know, we are the first to apply pre-training to adaptive learning in education and design a unified pre-training framework based on data analysis to model students' learning behaviors. We build a series of benchmarks for the selected tasks of \emph{Learning Recommendation}, \emph{Learning Resource Evaluation}, \emph{Knowledge Tracing}, and \emph{Dropout Prediction} from the three main research directions of adaptive learning to demonstrate the framework's effectiveness. The exploration reveals the high transferability and generalization to downstream tasks and the powerful capability of the pre-training method. On the one hand, it enlightens us about the specific roles of various learning elements and how they interact and complement each other to work together; on the other hand, the unified pre-training framework allows for better integration of diverse components, making the systematic application of adaptive learning possible. We believe that the research on pre-training in adaptive learning has broad prospects for both the fields of pedagogy and machine learning.

\ifCLASSOPTIONcaptionsoff
  \newpage
\fi

\bibliographystyle{IEEEtran}
\bibliography{reference.bib}

\begin{thebibliography}{10}
\providecommand{\url}[1]{#1}
\csname url@samestyle\endcsname
\providecommand{\newblock}{\relax}
\providecommand{\bibinfo}[2]{#2}
\providecommand{\BIBentrySTDinterwordspacing}{\spaceskip=0pt\relax}
\providecommand{\BIBentryALTinterwordstretchfactor}{4}
\providecommand{\BIBentryALTinterwordspacing}{\spaceskip=\fontdimen2\font plus
\BIBentryALTinterwordstretchfactor\fontdimen3\font minus
  \fontdimen4\font\relax}
\providecommand{\BIBforeignlanguage}[2]{{%
\expandafter\ifx\csname l@#1\endcsname\relax
\typeout{** WARNING: IEEEtran.bst: No hyphenation pattern has been}%
\typeout{** loaded for the language `#1'. Using the pattern for}%
\typeout{** the default language instead.}%
\else
\language=\csname l@#1\endcsname
\fi
#2}}
\providecommand{\BIBdecl}{\relax}
\BIBdecl

\bibitem{kaplan2021higher}
A.~Kaplan, \emph{Higher Education at the Crossroads of Disruption: The
  University of the 21st Century}.\hskip 1em plus 0.5em minus 0.4em\relax
  Emerald Group Publishing, 2021.

\bibitem{pan2017prerequisite}
L.~Pan, C.~Li, J.~Li, and J.~Tang, ``Prerequisite relation learning for
  concepts in moocs,'' in \emph{Proceedings of the 55th Annual Meeting of the
  Association for Computational Linguistics (Volume 1: Long Papers)}, 2017, pp.
  1447--1456.

\bibitem{yu2019course}
J.~Yu, C.~Wang, G.~Luo, L.~Hou, J.~Li, Z.~Liu, and J.~Tang, ``Course concept
  expansion in moocs with external knowledge and interactive game,'' in
  \emph{Proceedings of the 57th Conference of the Association for Computational
  Linguistics}, 2019, pp. 4292--4302.

\bibitem{piech2015deep}
C.~Piech, J.~Bassen, J.~Huang, S.~Ganguli, M.~Sahami, L.~J. Guibas, and
  J.~Sohl-Dickstein, ``Deep knowledge tracing,'' \emph{Advances in neural
  information processing systems}, vol.~28, pp. 505--513, 2015.

\bibitem{chen2018prerequisite}
P.~Chen, Y.~Lu, V.~W. Zheng, and Y.~Pian, ``Prerequisite-driven deep knowledge
  tracing,'' in \emph{2018 IEEE International Conference on Data Mining
  (ICDM)}.\hskip 1em plus 0.5em minus 0.4em\relax IEEE, 2018, pp. 39--48.

\bibitem{liu2019exploiting}
Q.~Liu, S.~Tong, C.~Liu, H.~Zhao, E.~Chen, H.~Ma, and S.~Wang, ``Exploiting
  cognitive structure for adaptive learning,'' in \emph{Proceedings of the 25th
  ACM SIGKDD International Conference on Knowledge Discovery \& Data Mining},
  2019, pp. 627--635.

\bibitem{bulathwela2020truelearn}
S.~Bulathwela, M.~Perez-Ortiz, E.~Yilmaz, and J.~Shawe-Taylor, ``Truelearn: A
  family of bayesian algorithms to match lifelong learners to open educational
  resources,'' in \emph{Proceedings of the AAAI Conference on Artificial
  Intelligence}, vol.~34, no.~01, 2020, pp. 565--573.

\bibitem{tkde2021recommendation}
Y.~Zhu, Q.~Lin, H.~Lu, K.~Shi, D.~Liu, J.~Chambua, S.~Wan, and Z.~Niu,
  ``Recommending learning objects through attentive heterogeneous graph
  convolution and operation-aware neural network,'' \emph{IEEE Transactions on
  Knowledge and Data Engineering}, 2021.

\bibitem{yu2021mooccubex}
J.~Yu, Y.~Wang, Q.~Zhong, G.~Luo, Y.~Mao, K.~Sun, W.~Feng, W.~Xu, S.~Cao,
  K.~Zeng \emph{et~al.}, ``Mooccubex: A large knowledge-centered repository for
  adaptive learning in moocs,'' in \emph{Proceedings of the 30th ACM
  International Conference on Information \& Knowledge Management}, 2021, pp.
  4643--4652.

\bibitem{feng2019understanding}
W.~Feng, J.~Tang, and T.~X. Liu, ``Understanding dropouts in moocs,'' in
  \emph{Proceedings of the AAAI Conference on Artificial Intelligence, (Vol 33
  No 01: AAAI-19, IAAI-19, EAAI-20)}, 2019.

\bibitem{ferguson2012learning}
R.~Ferguson, ``Learning analytics: drivers, developments and challenges,''
  \emph{International Journal of Technology Enhanced Learning}, vol.~4, no.
  5-6, pp. 304--317, 2012.

\bibitem{zhang2019hierarchical}
J.~Zhang, B.~Hao, B.~Chen, C.~Li, H.~Chen, and J.~Sun, ``Hierarchical
  reinforcement learning for course recommendation in moocs,'' in
  \emph{Proceedings of the AAAI Conference on Artificial Intelligence},
  vol.~33, no.~01, 2019, pp. 435--442.

\bibitem{gong2020sigir}
J.~Gong, S.~Wang, J.~Wang, W.~Feng, H.~Peng, J.~Tang, and P.~S. Yu,
  \emph{Attentional Graph Convolutional Networks for Knowledge Concept
  Recommendation in MOOCs in a Heterogeneous View}, 2020, p. 79–88.

\bibitem{fabbri2018tutorialbank}
A.~R. Fabbri, I.~Li, P.~Trairatvorakul, Y.~He, W.~Ting, R.~Tung,
  C.~Westerfield, and D.~Radev, ``Tutorialbank: A manually-collected corpus for
  prerequisite chains, survey extraction and resource recommendation,'' in
  \emph{Proceedings of the 56th Annual Meeting of the Association for
  Computational Linguistics (Volume 1: Long Papers)}, 2018, pp. 611--620.

\bibitem{shaffer2016tutorial}
D.~W. Shaffer, W.~Collier, and A.~R. Ruis, ``A tutorial on epistemic network
  analysis: Analyzing the structure of connections in cognitive, social, and
  interaction data,'' \emph{Journal of Learning Analytics}, vol.~3, no.~3, pp.
  9--45, 2016.

\bibitem{martin2020systematic}
F.~Martin, Y.~Chen, R.~L. Moore, and C.~D. Westine, ``Systematic review of
  adaptive learning research designs, context, strategies, and technologies
  from 2009 to 2018,'' \emph{Educational Technology Research and Development},
  vol.~68, no.~4, pp. 1903--1929, 2020.

\bibitem{Peters:2018}
M.~E. Peters, M.~Neumann, M.~Iyyer, M.~Gardner, C.~Clark, K.~Lee, and
  L.~Zettlemoyer, ``Deep contextualized word representations,'' in \emph{Proc.
  of NAACL}, 2018.

\bibitem{devlin2019bert}
J.~Devlin, M.-W. Chang, K.~Lee, and K.~Toutanova, ``Bert: Pre-training of deep
  bidirectional transformers for language understanding,'' in \emph{Proceedings
  of the 2019 Conference of the North American Chapter of the Association for
  Computational Linguistics: Human Language Technologies, Volume 1 (Long and
  Short Papers)}, 2019, pp. 4171--4186.

\bibitem{brown2020gpt3}
T.~B. Brown, B.~Mann, N.~Ryder, M.~Subbiah, J.~Kaplan, P.~Dhariwal,
  A.~Neelakantan, P.~Shyam, G.~Sastry, A.~Askell \emph{et~al.}, ``Language
  models are few-shot learners,'' \emph{arXiv preprint arXiv:2005.14165}, 2020.

\bibitem{raffel2019exploring}
C.~Raffel, N.~Shazeer, A.~Roberts, K.~Lee, S.~Narang, M.~Matena, Y.~Zhou,
  W.~Li, and P.~J. Liu, ``Exploring the limits of transfer learning with a
  unified text-to-text transformer,'' \emph{arXiv preprint arXiv:1910.10683},
  2019.

\bibitem{graf2007depth}
S.~Graf, S.~R. Viola, T.~Leo, and Kinshuk, ``In-depth analysis of the
  felder-silverman learning style dimensions,'' \emph{Journal of Research on
  Technology in Education}, vol.~40, no.~1, pp. 79--93, 2007.

\bibitem{ramesh2014learning}
A.~Ramesh, D.~Goldwasser, B.~Huang, H.~Daume~III, and L.~Getoor, ``Learning
  latent engagement patterns of students in online courses,'' in
  \emph{Twenty-Eighth AAAI Conference on Artificial Intelligence}, 2014.

\bibitem{wang2020neural}
F.~Wang, Q.~Liu, E.~Chen, Z.~Huang, Y.~Chen, Y.~Yin, Z.~Huang, and S.~Wang,
  ``Neural cognitive diagnosis for intelligent education systems,'' in
  \emph{Proceedings of the AAAI Conference on Artificial Intelligence},
  vol.~34, no.~04, 2020, pp. 6153--6161.

\bibitem{yu-etal-2020-mooccube}
\BIBentryALTinterwordspacing
J.~Yu, G.~Luo, T.~Xiao, Q.~Zhong, Y.~Wang, W.~Feng, J.~Luo, C.~Wang, L.~Hou,
  J.~Li, Z.~Liu, and J.~Tang, ``{MOOCC}ube: A large-scale data repository for
  {NLP} applications in {MOOC}s,'' in \emph{Proceedings of the 58th Annual
  Meeting of the Association for Computational Linguistics}.\hskip 1em plus
  0.5em minus 0.4em\relax Online: Association for Computational Linguistics,
  Jul. 2020, pp. 3135--3142. [Online]. Available:
  \url{https://www.aclweb.org/anthology/2020.acl-main.285}
\BIBentrySTDinterwordspacing

\bibitem{pardos2014affective}
Z.~A. Pardos, R.~S. Baker, M.~O. San~Pedro, S.~M. Gowda, and S.~M. Gowda,
  ``Affective states and state tests: Investigating how affect and engagement
  during the school year predict end-of-year learning outcomes.'' \emph{Journal
  of Learning Analytics}, vol.~1, no.~1, pp. 107--128, 2014.

\bibitem{pan2017course}
L.~Pan, X.~Wang, C.~Li, J.~Li, and J.~Tang, ``Course concept extraction in
  moocs via embedding-based graph propagation,'' in \emph{Proceedings of the
  Eighth International Joint Conference on Natural Language Processing (Volume
  1: Long Papers)}, vol.~1, 2017, pp. 875--884.

\bibitem{dalipi2018mooc}
F.~Dalipi, A.~S. Imran, and Z.~Kastrati, ``Mooc dropout prediction using
  machine learning techniques: Review and research challenges,'' in
  \emph{Proceedings of 2018 IEEE Global Engineering Education Conference
  (EDUCON)}.\hskip 1em plus 0.5em minus 0.4em\relax IEEE, 2018, pp. 1007--1014.

\bibitem{moreno2020temporal}
P.~M. Moreno-Marcos, P.~J. Munoz-Merino, J.~Maldonado-Mahauad,
  M.~Perez-Sanagustin, C.~Alario-Hoyos, and C.~D. Kloos, ``Temporal analysis
  for dropout prediction using self-regulated learning strategies in self-paced
  moocs,'' \emph{Computers \& Education}, vol. 145, p. 103728, 2020.

\bibitem{pre4recstrategy}
S.~Yang, Y.~Liu, L.~Chenyi, G.~Wang, H.~Tang, J.~Zhang, and C.~Miao, ``A
  pre-training strategy for recommendation,'' 10 2020.

\bibitem{shaffer2017epistemic}
D.~Shaffer and A.~Ruis, ``Epistemic network analysis: A worked example of
  theory-based learning analytics,'' \emph{Handbook of learning analytics},
  2017.

\bibitem{dosovitskiy2020image}
A.~Dosovitskiy, L.~Beyer, A.~Kolesnikov, D.~Weissenborn, X.~Zhai,
  T.~Unterthiner, M.~Dehghani, M.~Minderer, G.~Heigold, S.~Gelly \emph{et~al.},
  ``An image is worth 16x16 words: Transformers for image recognition at
  scale,'' \emph{arXiv preprint arXiv:2010.11929}, 2020.

\bibitem{ramesh2021zero}
A.~Ramesh, M.~Pavlov, G.~Goh, S.~Gray, C.~Voss, A.~Radford, M.~Chen, and
  I.~Sutskever, ``Zero-shot text-to-image generation,'' \emph{arXiv preprint
  arXiv:2102.12092}, 2021.

\bibitem{ding2021cogview}
M.~Ding, Z.~Yang, W.~Hong, W.~Zheng, C.~Zhou, D.~Yin, J.~Lin, X.~Zou, Z.~Shao,
  H.~Yang \emph{et~al.}, ``Cogview: Mastering text-to-image generation via
  transformers,'' \emph{arXiv preprint arXiv:2105.13290}, 2021.

\bibitem{qiu2020gcc}
J.~Qiu, Q.~Chen, Y.~Dong, J.~Zhang, H.~Yang, M.~Ding, K.~Wang, and J.~Tang,
  ``Gcc: Graph contrastive coding for graph neural network pre-training,'' in
  \emph{Proceedings of the 26th ACM SIGKDD International Conference on
  Knowledge Discovery \& Data Mining}, 2020, pp. 1150--1160.

\bibitem{roy2019inferring}
S.~Roy, M.~Madhyastha, S.~Lawrence, and V.~Rajan, ``Inferring concept
  prerequisite relations from online educational resources,'' in
  \emph{Proceedings of the AAAI Conference on Artificial Intelligence},
  vol.~33, 2019, pp. 9589--9594.

\bibitem{roberta}
Y.~Liu, M.~Ott, N.~Goyal, J.~Du, M.~Joshi, D.~Chen, O.~Levy, M.~Lewis,
  L.~Zettlemoyer, and V.~Stoyanov, ``Roberta: A robustly optimized bert
  pretraining approach,'' 2019.

\bibitem{attention}
A.~Vaswani, N.~Shazeer, N.~Parmar, J.~Uszkoreit, L.~Jones, A.~N. Gomez, L.~u.
  Kaiser, and I.~Polosukhin, ``Attention is all you need,'' in \emph{Advances
  in Neural Information Processing Systems}, vol.~30, 2017.

\bibitem{word2vec}
T.~Mikolov, K.~Chen, G.~Corrado, and J.~Dean, ``Efficient estimation of word
  representations in vector space,'' 2013.

\bibitem{sun2019bert4rec}
F.~Sun, J.~Liu, J.~Wu, C.~Pei, X.~Lin, W.~Ou, and P.~Jiang, ``Bert4rec:
  Sequential recommendation with bidirectional encoder representations from
  transformer,'' in \emph{Proceedings of the 28th ACM international conference
  on information and knowledge management}, 2019, pp. 1441--1450.

\bibitem{meantime}
S.~M. Cho, E.~Park, and S.~Yoo, ``Meantime: Mixture of attention mechanisms
  with multi-temporal embeddings for sequential recommendation,'' in
  \emph{Fourteenth {ACM} Conference on Recommender Systems}.\hskip 1em plus
  0.5em minus 0.4em\relax ACM, sep 2020.

\bibitem{he2017neural}
X.~He, L.~Liao, H.~Zhang, L.~Nie, X.~Hu, and T.-S. Chua, ``Neural collaborative
  filtering,'' in \emph{Proceedings of the 26th international conference on
  world wide web}, 2017, pp. 173--182.

\bibitem{kang2018self}
W.-C. Kang and J.~McAuley, ``Self-attentive sequential recommendation,'' in
  \emph{Proceedings of 2018 IEEE International Conference on Data Mining
  (ICDM)}.\hskip 1em plus 0.5em minus 0.4em\relax IEEE, 2018, pp. 197--206.

\bibitem{tang2018personalized}
J.~Tang and K.~Wang, ``Personalized top-n sequential recommendation via
  convolutional sequence embedding,'' in \emph{Proceedings of the Eleventh ACM
  International Conference on Web Search and Data Mining}, 2018, pp. 565--573.

\bibitem{hidasi2015session}
B.~Hidasi, A.~Karatzoglou, L.~Baltrunas, and D.~Tikk, ``Session-based
  recommendations with recurrent neural networks,'' \emph{arXiv preprint
  arXiv:1511.06939}, 2015.

\bibitem{tyler1967changing}
R.~W. Tyler, ``Changing concepts of educational evaluation.'' 1967.

\bibitem{chen2014investigating}
Y.~Chen, ``Investigating moocs through blog mining,'' \emph{The International
  Review of Research in Open and Distributed Learning}, vol.~15, no.~2, 2014.

\bibitem{reckase2009multidimensional}
M.~D. Reckase, ``Multidimensional item response theory models,'' in
  \emph{Multidimensional item response theory}.\hskip 1em plus 0.5em minus
  0.4em\relax Springer, 2009, pp. 79--112.

\bibitem{drasgow1990item}
F.~Drasgow and C.~L. Hulin, ``Item response theory.'' 1990.

\bibitem{shah2020year}
D.~Shah, ``The second year of the mooc: A review of mooc stats and trends in
  2020,'' \emph{Class Central’s MOOC Report}, 2020.

\end{thebibliography}

\begin{IEEEbiography}[{\includegraphics[width=0.9in,height=1in,clip,keepaspectratio]{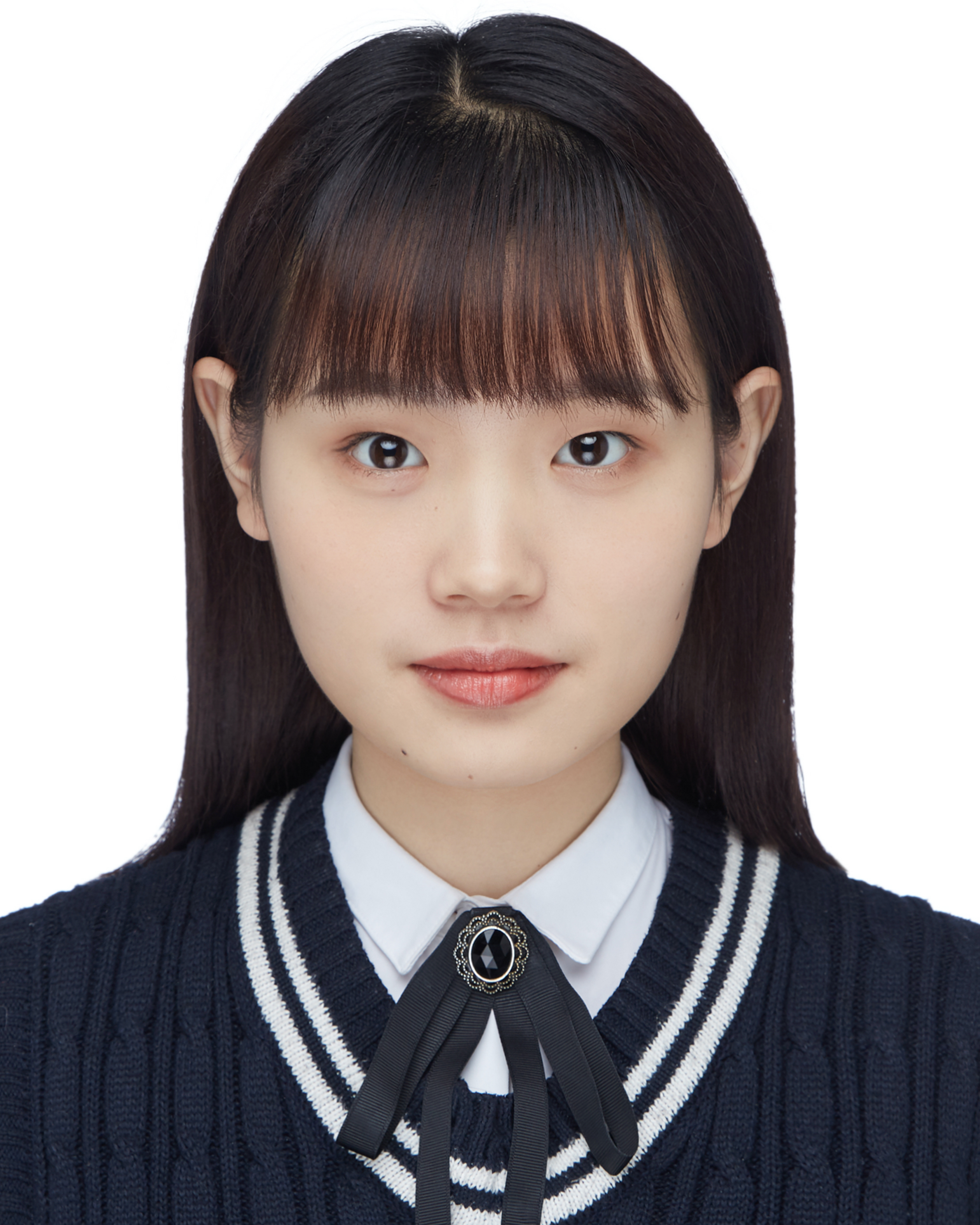}}]{Qingyang Zhong} is a master student in the Department of Computer Science and Technology, Tsinghua University. She got her bachelor's degree from the Department of Automation, Tsinghua University. Her research interests include intelligent education systems and knowledge discovery.
\end{IEEEbiography} %

\begin{IEEEbiography}[{\includegraphics[width=0.92in,height=1in,clip,keepaspectratio]{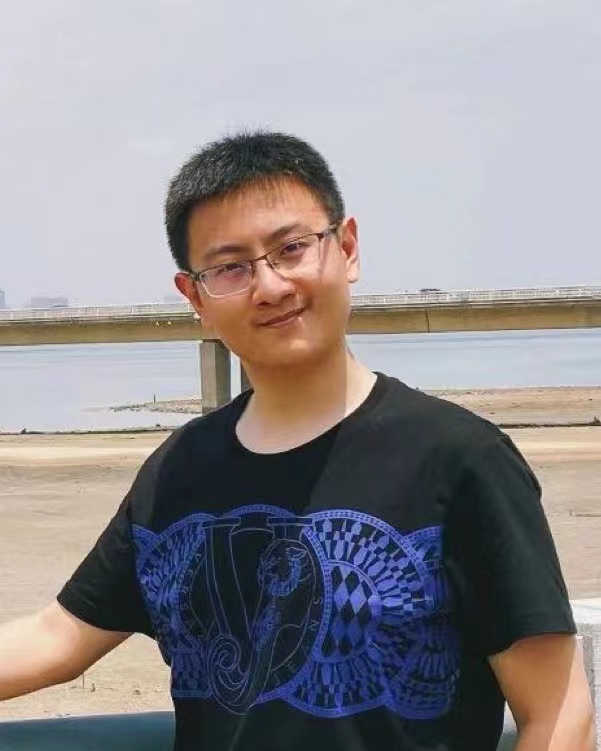}}]{Jifan Yu} is a Ph.D candidate in the Department of Computer Science and Technology, Tsinghua University. He got his bachelor degree in Beihang University. His research interests include AI-driven MOOCs, Knowledge-grounded generation and low-resource information extraction.
\end{IEEEbiography}

\begin{IEEEbiography}[{\includegraphics[width=0.92in,height=1.1in,clip,keepaspectratio]{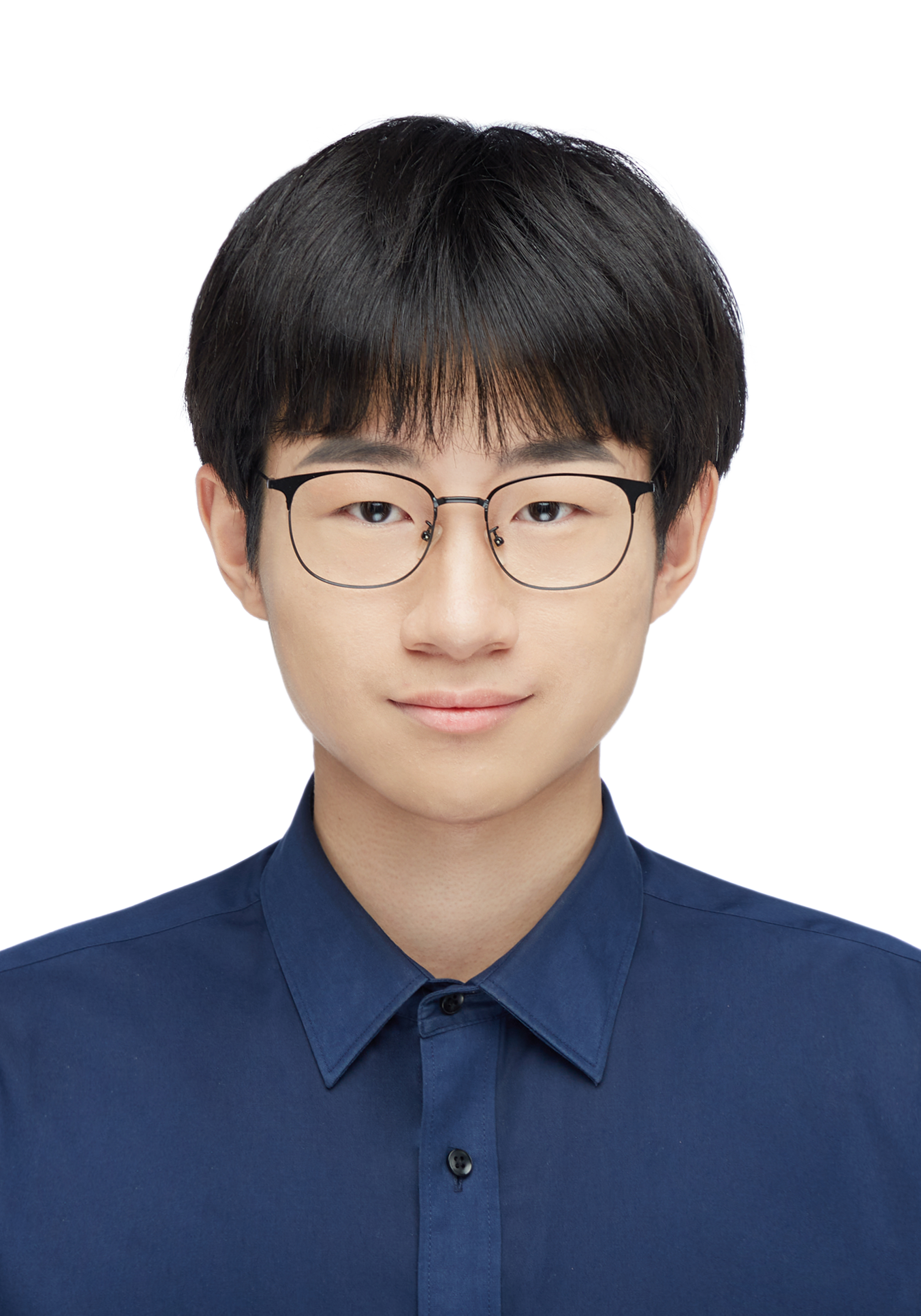}}]{Zheyuan Zhang} is an undergraduate student in Xinya College, Tsinghua University. He is about to study for a master's degree in the Department of Computer Science and Technology in Tsinghua University in 2022. His research interests include knowledge engineering and natural language processing.  
\end{IEEEbiography}

\begin{IEEEbiography}[{\includegraphics[width=1in,height=1.15in,clip,keepaspectratio]{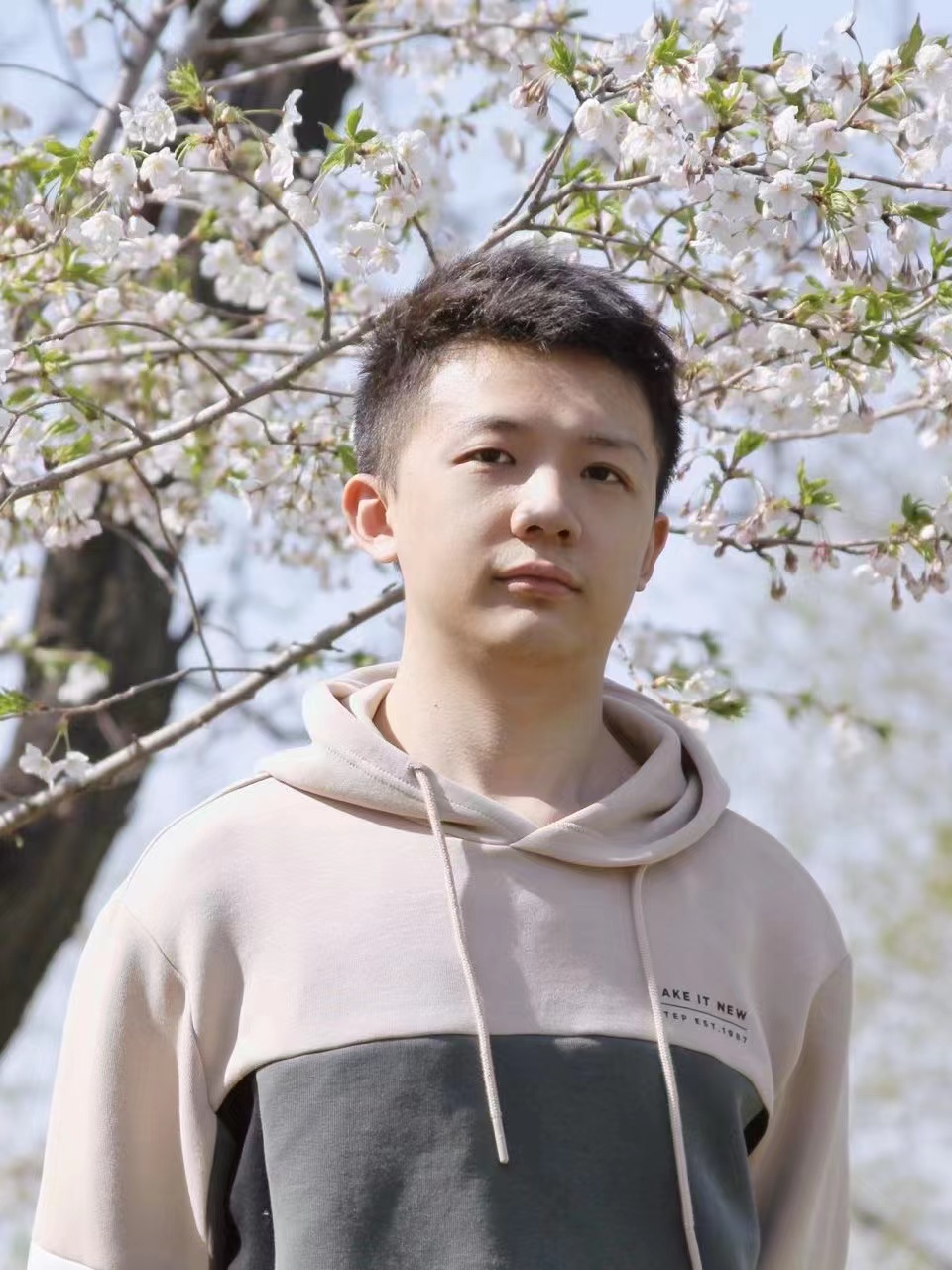}}]{Yiming Mao} is working in Tencent Group, He got his master degree from Tsinghua University in 2021. He also got his bachelor degree in Computer Science and Technology from Tsinghua University. His interests include graph learning, recommendation system and adapitve learning.
\end{IEEEbiography}

\begin{IEEEbiography}[{\includegraphics[width=1in,height=1.05in,clip,keepaspectratio]{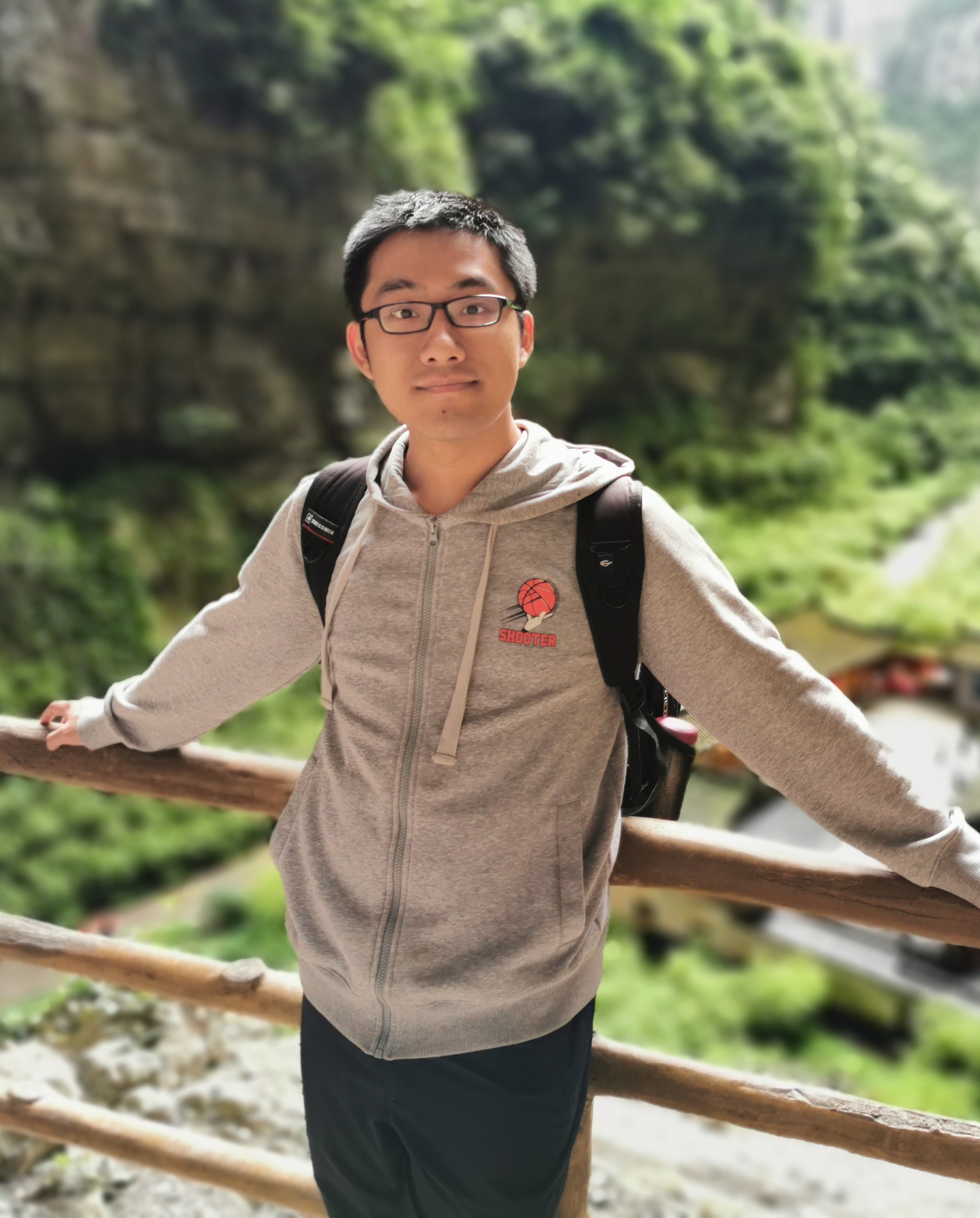}}]{Yuquan Wang} is a graduate student in the Department of Computer Science and Technology in Tsinghua University. He got his bechelor degree in Shanghai Jiao Tong University. His research interests include natural language processing and machine learning.
\end{IEEEbiography}

\begin{IEEEbiography}[{\includegraphics[width=1in,height=1.3in,clip,keepaspectratio]{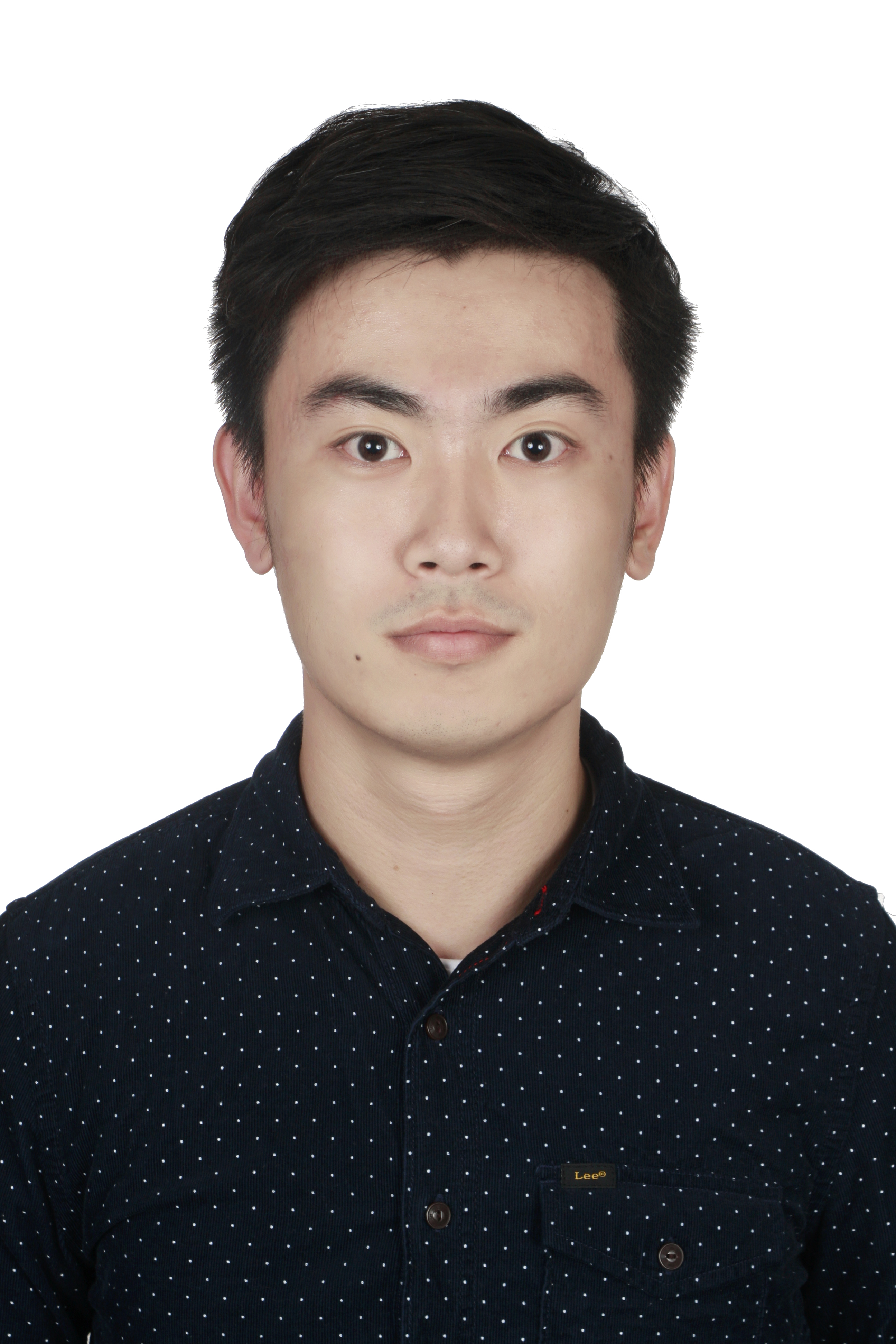}}]{Yankai Lin} is a researcher in WeChat AI, Tencent Technology Company Limited. He obtained his Ph.D. and bachelor's degrees from the Department of Computer Science and Technology at Tsinghua University in 2019 and 2014. He has published many papers on top international conferences, such as ACL and EMNLP. His research interests are in natural language processing.
\end{IEEEbiography}

\begin{IEEEbiography}[{\includegraphics[width=1in,height=1.1in,clip,keepaspectratio]{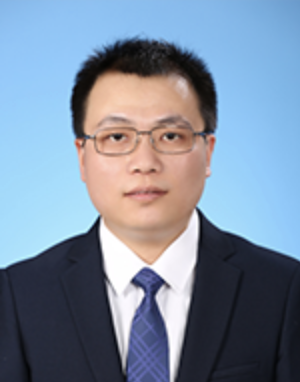}}]{Lei Hou} is an assistant researcher in the Department of Computer Science and Technology at Tsinghua University. He obtained his Ph.D. degrees from Tsinghua University in 2016. He has published many research papers on top international conferences, such as ACL, EMNLP, AAAI, etc. His research interests include knowledge graph construction and application, news and user-generated content mining.
\end{IEEEbiography}

\begin{IEEEbiography}[{\includegraphics[width=1in,height=1.1in,clip,keepaspectratio]{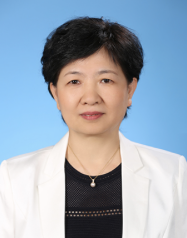}}]{Juanzi Li} is a Professor of the Department of Computer Science and Technology at Tsinghua University. She received her Ph.D. degrees from Tsinghua University in 2000. Her research interests include knowledge engineering and semantic web, text and social network mining. She has published over 100 research papers in top international journals and conferences such as ACL, EMNLP, AAAI, etc.
\end{IEEEbiography}

\begin{IEEEbiography}[{\includegraphics[width=0.9in,height=0.9in,clip,keepaspectratio]{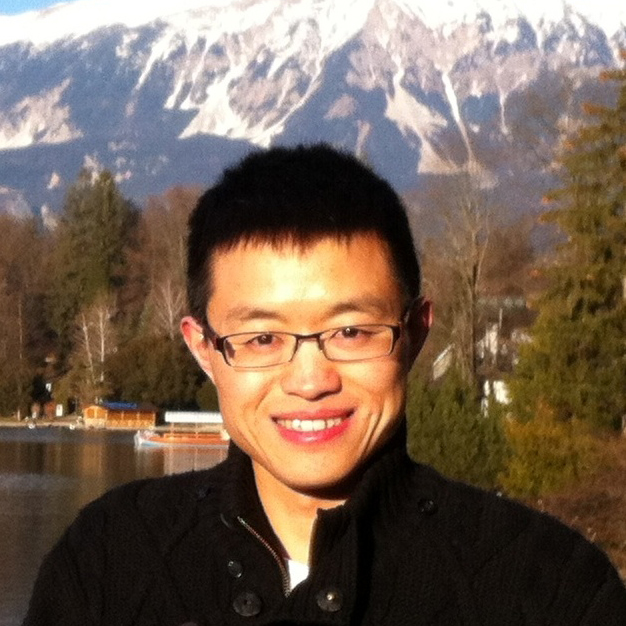}}]{Jie Tang} is a Professor and the Associate Chair of the Department of Computer Science and Technology at Tsinghua University. He is a Fellow of the ACM and IEEE. His research interests include artificial general intelligence, data mining, social networks, machine learning and knowledge graph. He has published over 300 research papers in top international journals and conferences and was honored with the SIGKDD Test-of-Time Award. 
\end{IEEEbiography}

\end{document}